\documentclass[headings=standardclasses]{scrartcl}

\usepackage{geometry}
\usepackage[utf8]{inputenc}
\usepackage[T1]{fontenc}
\usepackage[onehalfspacing]{setspace}
\usepackage{charter}
\usepackage[charter]{mathdesign}
\usepackage{microtype}
\usepackage{amsmath}
\usepackage{amsthm}
\usepackage{tabulary}
\usepackage{longtable}
\usepackage{booktabs} 
\usepackage{diagbox}
\usepackage{ragged2e}

\usepackage{siunitx}
\usepackage{bm}
\usepackage[pdftex]{graphicx}
\usepackage{rotating}
\usepackage{pdfpages}
\usepackage{pdflscape}
\usepackage{color}
\usepackage{array}
\usepackage[font=bf]{caption}
\usepackage{subcaption}
\usepackage{pgfplots}
\usepackage{pgfplotstable}
\usepackage{float}
\usepackage[title]{appendix}
\pgfplotsset{compat=1.9}
\usepackage{url}
\usepackage[section]{placeins}
\usepackage{natbib}
\usepackage{hyperref}
\hypersetup{
	colorlinks   = true,    
	urlcolor     = blue,    
	linkcolor    = blue,    
	citecolor    = blue      
}

\usepackage[boxed]{algorithm2e}
\SetKwFunction{KwInit}{Initialisation}
\SetKwComment{Comment}{\% }{}

\makeatletter
\renewcommand*\env@matrix[1][c]{\hskip -\arraycolsep
  \let\@ifnextchar\new@ifnextchar
  \array{*\c@MaxMatrixCols #1}}
\makeatother

\newcommand\blfootnote[1]{%
  \begingroup
  \renewcommand\thefootnote{}\footnote{#1}%
  \addtocounter{footnote}{-1}%
  \endgroup
}

\usepackage{xcolor}
\usepackage{soul}

\usepackage{verbatim}

\begin{document}
\thispagestyle{empty}
\begin{spacing}{1.2}
\begin{flushleft}
\huge \textbf{Fast Bayesian Estimation of Spatial Count Data Models} \\
\vspace{\baselineskip}
\normalsize
15 October 2020 \\
\vspace{\baselineskip}
\textsc{Prateek Bansal}\textsuperscript{*} (corresponding author)\blfootnote{Postal address: 610, Skempton Building, 
South Kensington Campus, Imperial College London, Tel. +44-7307278771.}\\
Transport Strategy Centre, Department of Civil and Environmental Engineering\\
Imperial College London, UK \\
prateek.bansal@imperial.ac.uk \\
\vspace{\baselineskip}
\textsc{Rico Krueger}\textsuperscript{*} \\
Transport and Mobility Laboratory \\
Ecole Polytechnique F\'{e}d\'{e}rale de Lausanne, Switzerland \\
rico.krueger@epfl.ch \\
\vspace{\baselineskip}
\textsc{Daniel J. Graham}\\
Transport Strategy Centre, Department of Civil and Environmental Engineering\\
Imperial College London, UK \\
d.j.graham@imperial.ac.uk  \\
\vspace{\baselineskip}
\textsuperscript{*} Equal contribution.
\end{flushleft}
\end{spacing}

\blfootnote{Additional simulation results will appear as annexes in the electronic version.} 

\newpage
\thispagestyle{empty}
\section*{Abstract}

Spatial count data models are used to explain and predict the frequency of phenomena such as traffic accidents in geographically distinct entities such as census tracts or road segments. These models are typically estimated using Bayesian Markov chain Monte Carlo (MCMC) simulation methods, which, however, are computationally expensive and do not scale well to large datasets. Variational Bayes (VB), a method from machine learning, addresses the shortcomings of MCMC by casting Bayesian estimation as an optimisation problem instead of a simulation problem. Considering all these advantages of VB, a VB method is derived for posterior inference in negative binomial models with unobserved parameter heterogeneity and spatial dependence.  P{\'o}lya-Gamma augmentation is used to deal with the non-conjugacy of the negative binomial likelihood and an integrated non-factorised specification of the variational distribution is adopted to capture posterior dependencies. The benefits of the proposed approach are demonstrated in a Monte Carlo study and an empirical application on estimating youth pedestrian injury counts in census tracts of New York City. The VB approach is around 45 to 50 times faster than MCMC on a regular eight-core processor in a simulation and an empirical study, while offering similar estimation and predictive accuracy. Conditional on the availability of computational resources, the embarrassingly parallel architecture of the proposed VB method can be exploited to further accelerate its estimation by up to 20 times.
\\
\\
\textit{Keywords:} Variational Bayes; spatial count data; negative binomial regression; P{\'o}lya-Gamma data augmentation; accident analysis.

\newpage

\section{Introduction}

Spatial count data models are widely used in disciplines such as ecology, epidemiology, geography, regional science as well as transportation planning and engineering to explain and predict non-negative integer-valued outcome variables such as species and disease counts, patenting and innovation activities as well as crime and accident rates in geographically distinct entities such as local government areas, census tracts or traffic analysis zones \citep[e.g.][]{acs2002patents, dormann2007methods, glaser2017review, marshall1991review, ver2018spatial, wakefield2007disease}. 

Models of spatial count data typically pivot on Poisson lognormal and negative binomial regressions, in which the spatial arrangement of the investigated units is explicitly specified. These models generally consider two types of spatial effects, namely \emph{spatial heterogeneity} and \emph{spatial dependence} \citep{simoes2016spatial}. While \emph{spatial heterogeneity} accounts for the spatially-varying effect of covariates on the dependent variable, \emph{spatial dependence} captures the systematic correlation across neighbouring spatial units. In spatial count data models, unobserved spatial heterogeneity is operationalised through the inclusion of random link function parameters \citep{mannering2016unobserved}; spatial dependence can be represented through different variants of autoregressive specifications including the spatial and conditional autoregressive and matrix exponential spatial specifications \citep{whittle1954stationary, besag1974spatial, lesage2007matrix}. Ignoring these spatial effects may result in biased parameter estimates and inaccurate inference due to higher type-I error \citep{anselin2013spatial, dormann2007effects, dormann2007methods}. However, accounting for spatial heterogeneity and dependence also renders the estimation of spatial count data models computationally expensive.

Spatial count data models are predominantly estimated using Markov Chain Monte Carlo (MCMC) methods \citep{banerjee2014hierarchical, haining2020regression}, aside from few exceptions which rely on maximum likelihood estimation \citep{castro2012latent,narayanamoorthy2013accommodating}. 
MCMC methods guarantee asymptotically exact inference, but succumb to three important limitations, namely computationally intensive estimation, high storage costs for the posterior draws, and difficulties in assessing convergence \citep{bansal2020bayesian}. 
Furthermore, state-of-practice Gibbs samplers for spatial count data models also include Metropolis-Hastings steps to sample from high-dimensional conditional distributions, since conjugate priors for the parameters of Poisson lognormal and negative binomial regressions are not known. Sampling via the Metropolis-Hastings algorithm suffers from a variety of inefficiencies including insufficient exploration of the posterior of interest and serial correlation, if it is not tuned well \citep{rossi2012bayesian}.

To address the bottlenecks of MCMC in the estimation of spatial econometric models, \citet{bivand2014approximate} propose the integrated nested Laplace approximation (INLA) method, under which the model parameters are first segregated into hyper-parameters and latent variables. Then, a discrete distribution is specified on the hyper-parameters using a multi-dimensional grid, and the posterior distribution of the latent variables is approximated via Laplace's method. This analytical approximation comes at the cost of the assumption that conditional on the hyper-parameters, the latent variables are normally distributed. INLA reduces the estimation times of typical spatial econometric models from hours to minutes, but the conditional normality assumption restricts the flexibility of the posterior approximation \citep{han2013integrated}. 

In machine learning and computational statistics, variational Bayes (VB) methods have also emerged as a promising alternative to MCMC for the estimation of complex econometric models \citep{bansal2020bayesian, blei2017variational,braun2010variational,jordan1999introduction, tan2013variational}. Whilst MCMC treats Bayesian inference as a simulation problem, in which the posterior distribution of interest is approximated through samples from a Markov chain, VB recasts Bayesian inference into an optimisation problem, which consists of minimising the probability distance between an approximating variational distribution and the targeted posterior distribution. Translating Bayesian inference into an optimisation problem accelerates estimation, admits a straightforward assessment of convergence and alleviates storage requirements. 

VB methods have been introduced for the estimation of non-spatial count data models and of linear spatial models. Yet, no VB method exists for the estimation of spatial count data models. Several studies present VB methods for variants of count data models, but none of the proposed approaches accounts for spatial dependencies between units \citep{klami2015polya, luts2015variational, tan2013variational, zhou2012lognormal}. \citet{kabisa2016online}, \citet{ren2011variational} and \citet{wu2018fast} devise VB methods for the estimation of models with spatial dependence; however, the proposed methods are limited to linear models with continuous outcome variables. 

In this paper, we propose a VB method for the fast estimation of a spatial count data model, which accommodates both spatial heterogeneity and dependence. To be specific, we consider a negative binomial (NB) model with random link function parameters and a matrix exponential spatial specification of spatial dependence \citep{lesage2007matrix}. To address the non-conjugacy of the NB model, we also adopt the P{\'o}lya-Gamma data augmentation (PGDA) technique in the proposed inference method. PDGA introduces auxiliary latent variables into the models. Conditional on these variables, the NB likelihood of the observed counts is translated into a heteroskedastic Gaussian likelihood, which admits closed-form conjugate posterior updates for nearly all model parameters. Only a few studies employ the PGDA technique in VB estimation \citep{durante2019conditionally, klami2015polya, park2016variational, wenzel2019efficient, zhou2012lognormal}. 

We first derive a mean-field variational Bayes (MFVB) method, which posits a factorised representation of the joint variational distributions, for the P{\'o}lya-Gamma-augmented spatial NB model. MFVB is the workhorse approach for the specification of the approximating variational distribution in VB inference. However, in the current application, the mean-field assumption oversimplifies posterior dependencies and leads to a high bias in the recovery of the spatial model parameters. Alternatively, the variational distribution can be specified according to the integrated non-factorised variational Bayes \citep[INFVB;][]{han2013integrated} approach, which generalises INLA by relaxing the conditional normality assumption. Motivated by the superior finite sample properties of INFVB for linear spatial models, we devise an INFVB method to allow for richer representations of relevant posterior dependencies in the considered spatial count data model. We benchmark the performance of INFVB against MCMC using simulated data and real data on youth pedestrian injury counts in New York City. The results indicate that INFVB is able to emulate the performance of MCMC in terms of posterior recovery and in-sample predictive accuracy. Furthermore, the embarrassingly parallel nature of the proposed INFVB algorithm makes INFVB substantially faster than MCMC, which, in turn, suggests that INFVB is scalable to large datasets of spatial counts. 

We organise the remainder of the paper as follows. In the subsequent section, we formulate the considered spatial negative binomial model, and in Section \ref{sec:est}, we derive MCMC and VB estimators for the model. In Section \ref{sec:sim}, we benchmark computational efficiency and finite sample properties of the proposed estimators in a Monte Carlo study. Section \ref{sec:app} further compares VB and MCMC in estimating youth pedestrian injury counts in the census tracts of New York City. The findings of this empirical application corroborate the insights derived from the simulation study. Conclusions and avenues for future research are presented in Section \ref{sec:conc}. 

\section{Model formulation} \label{sec:model}

Let $y_{i}$ denote the non-negative integer-valued outcome variable observed for spatial unit $i \in \{1,\dots,N\}$. We assume that $y_{i}$ is drawn from a negative binomial (NB) distribution with probability parameter $p_{i}$ and shape parameter $r$. We model $p_{i}$, using a logit link function, which depends on predictors $\bm{M}_{i}$ with fixed parameters $\bm{\gamma}$, predictors $\bm{X}_{i}$ with spatially-varying parameters $\bm{\beta}_{i}$ and a spatial random effect $\phi_{i}$. The resulting NB model is succinctly summarised below:
\begin{align}
& y_{i} \sim \text{NB}(r,p_i), & & i = 1,\dots,N \\
& p_i = \frac{\exp(\psi_i)}{1+\exp(\psi_i)}, & & i = 1,\dots,N\\
& \psi_i = \bm{M}_{i}^{\top} \bm{\gamma} + \bm{X}_{i}^{\top} \bm{\beta}_{i} + \phi_{i}.  & & i = 1,\dots,N   \label{eq_psi} 
\end{align} 

\subsection{Spatial heterogeneity and dependence} 
To accommodate spatial heterogeneity in the model, i.e. to allow for spatially varying effects of $\bm{X}_{i}$ on $y_{i}$, we place a multivariate Gaussian prior on $\bm{\beta}_{i}$ with mean $\bm{\mu}$ and covariance matrix $\bm{\Sigma}$. Furthermore, we apply the matrix exponential spatial specification \citep[MESS;][]{lesage2007matrix} to the random effect vector $\bm{\phi} = \left ( \phi_{1}, \ldots, \phi_{N} \right)^{\top}$ to capture spatial dependence between units. MESS is an attractive representation of spatial error dependence, as it implies a simple likelihood. Alternative specifications spatial dependence such as the spatial and conditional autoregressive ones, are similar to MESS with the key difference that MESS assumes an exponential decay instead of a geometric decay of spatial correlation \citep[see][for a detailed comparison]{strauss2017matrix}. The spatial aspects of the considered model are succinctly restated below:        
\begin{align}
& \bm{\beta}_{i}  \sim \text{Normal}(\bm{\mu}, \bm{\Sigma}), & & i = 1,\dots,N \\
& \bm{S}\bm{\phi} = \exp(\tau \bm{W})\bm{\phi} =  \bm{\epsilon}, \\
& \bm{\epsilon} \sim \text{Normal}(0,\sigma^2 \bm{I}_{N}).
\end{align} 
Here, $\bm{W}$ is a row-normalised spatial weight matrix, $\tau$ is the spatial association parameter, $\epsilon$ is a homoskedastic Gaussian error with scale $\sigma$,  and $\bm{I}_{N}$ is an identity matrix of size $N \times N$. $\exp(\tau \bm{W})$ is a matrix of size $N \times N$ given by a power series: $\sum_{k=0}^{\infty} \frac{\tau^{k}}{k!}\bm{W}^{k}$, where $\bm{W}^{0}$ is an identity matrix. We compute this matrix exponential using the Pade approximation \citep{al2010new}. 

\subsection{Model likelihood} 

Suppose that there are $Q$ fixed parameters and $K$ random parameters. Equation \ref{eq_psi} can be rewritten in vector form as follows: 
\begin{equation}
    \bm{\psi} = \bm{M} \bm{\gamma} + \bm{X} \bm{\beta} + \bm{\phi}, \\
\end{equation}
where
\begin{equation*}
 \bm{\psi} = \begin{bmatrix} 
   \psi_{1} \\
   \vdots\\
   \psi_{N} \\
    \end{bmatrix}_{N \times 1},\; 
 \bm{M} = \begin{bmatrix} 
   \bm{M}_{1}^{\top} \\
   \vdots\\
    \bm{M}_{N}^{\top}\\
    \end{bmatrix}_{N \times Q},\; 
\bm{X} = \begin{bmatrix} 
   \bm{X}_{1}^{\top} & \dots & 0 \\
    \vdots & \ddots &  \vdots\\
    0 &    \dots    &  \bm{X}_{N}^{\top} 
    \end{bmatrix}_{N \times NK},\; 
 \bm{\beta} = \begin{bmatrix} 
   \bm{\beta}_{1} \\
   \vdots\\
    \bm{\beta}_{N}\\
    \end{bmatrix}_{NK \times 1} .
\end{equation*}
Furthermore, note that $\Tilde{\bm{\Omega}} = \frac{\bm{S}^{\top} \bm{S}}{\sigma^2}$ and  $\text{det}(\bm{S}) = 1$, and thus, $\text{det}(\Tilde{\bm{\Omega}}) =(\sigma^2)^{-N}$ \citep{wu2018fast}. Consequently, the likelihood of the model is: 
\begin{equation} \label{eq:like}
\begin{split}
    P(\bm{y} \lvert r, \bm{\gamma}, \bm{\mu}, \bm{\Sigma},\sigma^{2}, \tau)  & = P(\bm{y}  \lvert r, \bm{\psi}) P(\bm{\psi} \lvert \bm{\gamma}, \bm{\beta},\bm{\phi}) P(\bm{\phi} \lvert \sigma^{2}, \tau) P(\bm{\beta} \lvert \bm{\mu}, \bm{\Sigma}), \\
    & = P(\bm{y}  \lvert r, \bm{\psi}) P(\bm{\psi} \lvert \bm{\gamma}, \bm{\beta},\sigma^{2}, \tau) P(\bm{\beta} \lvert \bm{\mu}, \bm{\Sigma}),  
\end{split}
\end{equation}
where 
\begin{equation} 
\begin{split}
    & P(\bm{y}  \lvert r, \bm{\psi}) =  \prod_{i=1}^N \frac{\Gamma(y_{i}+r)}{\Gamma(r) y_{i}!} \frac{\exp(\psi_{i})^{y_{i}}}{[1+\exp(\psi_{i})]^{r+y_{i}}},  \\
    & P(\bm{\psi} \lvert \bm{\gamma}, \bm{\beta},\sigma^{2}, \tau)  =  (2\pi\sigma^2)^{-\frac{N}{2}} \exp\left( -\frac{ [\bm{\psi} - \bm{M} \bm{\gamma} - \bm{X} \bm{\beta}]^{\top} \Tilde{\bm{\Omega}} [\bm{\psi} - \bm{M} \bm{\gamma} - \bm{X} \bm{\beta}]}{2} \right) , \\
    & P(\bm{\beta} \lvert \bm{\mu}, \bm{\Sigma}) = [2\pi \text{det}(\bm{\Sigma})]^{-\frac{N}{2}} \prod_{i=1}^{N} \exp \left( -\frac{1}{2}[\bm{\beta}_{i} - \bm{\mu}]^{\top} \bm{\Sigma}^{-1}[\bm{\beta}_{i} -  \bm{\mu}]\right).
\end{split}
\end{equation}

\section{Model estimation} \label{sec:est}

\subsection{P{\'o}lya-Gamma data augmentation} 

Conjugate priors for the parameters of the NB model are generally unknown. As a consequence, the conditional distributions of the link function parameters and the shape parameter do not constitute known distributions, and no closed-form updates for the respective model parameters exist \citep{klami2015polya,zhou2012lognormal}. To address this issue, \citet{polson2013bayesian} suggest to introduce P{\'o}lya-Gamma-distributed auxiliary variables $\omega_{i} \sim \text{PG}(y_{i} + r,0), i \in \{1,2,\dots,N\}$ into the model. Using the identity derived by \citet{polson2013bayesian}, $P(\bm{y}  \lvert r, \bm{\psi})$ can be written as:      
\begin{equation} \label{eq:polya}
        P(\bm{y}  \lvert r, \bm{\psi})
         = \prod_{i=1}^{N}\frac{\Gamma{(y_i+r)}}{\Gamma{(r)}y_i!} 2^{-(r+y_i)}\exp\left(\frac{(y_i-r)\psi_i}{2}\right) \mathbb{E}_{\omega_i}\left[\exp\left(\frac{-\omega_i \psi_i^2}{2}\right)\right]. 
\end{equation}
Furthermore, conditional on the auxiliary variables $\bm{\omega}$, equation \ref{eq:polya} can be restated as:  
\begin{equation}
    \begin{split}
         P(\bm{y} \lvert \bm{\psi},r,\bm{\omega}) & \propto \prod_{i=1}^{N} \exp\left(- \frac{\omega_{i}}{2}\left[\psi_{i} - \frac{y_{i}-r}{2\omega_{i}} \right]^2\right), \\
        P(\bm{y} \lvert \bm{\psi},r,\bm{\omega}) & \propto \exp\left( -\frac{1}{2}[\bm{\psi} - \bm{Z}]^{\top} \bm{\Omega} [\bm{\psi} - \bm{Z}]\right), 
    \end{split}
\end{equation}
where 
\begin{equation*}
 \bm{Z} = \begin{bmatrix} 
    \frac{y_{1}-r}{2\omega_{1}}\\
   \vdots\\
    \frac{y_{N}-r}{2\omega_{N}}\\
    \end{bmatrix}_{N \times 1}, \;
\bm{\Omega} = \begin{bmatrix} 
   \omega_{1} & \dots & 0 \\
    \vdots & \ddots &  \vdots\\
    0 &    \dots    &  \omega_{N} 
    \end{bmatrix}_{N \times N},
\end{equation*}
\begin{equation}\label{eq:aug}
\bm{Z} = \bm{\psi} + \bm{\alpha} =  \bm{M} \bm{\gamma} + \bm{X} \bm{\beta} + \bm{\phi} + \bm{\alpha}, \quad \bm{\alpha} \sim \text{Normal}(\bm{0},\bm{\Omega}^{-1}). 
\end{equation}
The main result of P{\'o}lya-Gamma data augmentation is that conditional on $r$ and $\bm{\omega}$, the likelihood of the observed counts is converted into a heteroskedastic Gaussian likelihood, which considers $\bm{Z}$ as outcome variable. As a consequence, we are able to obtain closed-form updates for the link function parameters and the shape parameter of the spatial NB model. 

\subsection{Prior specification and augmented likelihood} 

Prior distributions on latent variables are succinctly stated below: 
\begin{alignat*}{3}
& \bm{\mu} \sim \text{Normal}(\bm{\zeta}_{\bm{\mu}}, \bm{\Delta}_{\bm{\mu}}), \quad \quad \quad && \bm{\gamma} \sim \text{Normal}(\bm{\zeta}_{\bm{\gamma}}, \bm{\Delta}_{\bm{\gamma}}), \quad \quad \quad  && \tau \sim \text{Normal}(\zeta_{\tau}, \sigma_{\tau}^2), \\
& \sigma^{-2} \sim \text{Gamma} (b_{\sigma^2},c_{\sigma^2}), \quad \quad \quad && r \lvert h \sim \text{Gamma}(r_{0},h), \quad \quad \quad && h \sim \text{Gamma}(b_{0},c_{0}), \\
& \{a_{k}\}_{k=1}^{K}  \sim \text{Gamma}\left( s, \eta_{k}\right), \quad \quad \quad  && \bm{\Sigma} \lvert \bm{a} \sim \text{IW}\left(\rho, \bm{B} \right), 
\end{alignat*}
where 
$\rho = \nu + K - 1$, 
$\bm{a} = \begin{bmatrix} a_{1} & \dots & a_{K} \end{bmatrix}^{\top}$,
$\bm{B} = 2\nu \text{diag}(\bm{a})$, 
$s = \frac{1}{2}$ and
$\eta_{k} = A_{k}^{-2}$. 
We specify Huang's half-t prior on the covariance matrix of random parameters $\bm{\Sigma}$ by introducing $\bm{a}$ \citep{huang2013simple}. Here  $\{\bm{\zeta}_{\bm{\mu}}, \bm{\Delta}_{\bm{\mu}},\bm{\zeta}_{\bm{\gamma}}, \bm{\Delta}_{\bm{\gamma}}, \zeta_{\tau}, \sigma_{\tau}^2,  b_{\sigma^2},c_{\sigma^2}, r_{0}, b_{0},c_{0}, \nu, \{A_{k}\}_{k=1}^K \}$ is a set of hyper-parameters and $\bm{\Theta} = \left\{\bm{\phi}, \bm{\gamma}, \bm{\beta}, \bm{\mu}, \bm{a},\bm{\Sigma}, \sigma^{2}, \bm{\omega}, r, h, \tau \right\}$ is a set of latent variables. The joint distribution of latent and observed variables is:    
\begin{equation}
    \begin{split}
        P(\bm{y} , \bm{\Theta}) & =  P(\bm{Z} \lvert r,\bm{\omega}, \bm{\gamma}, \bm{\beta}, \bm{\phi}) P(\bm{\phi} \lvert \sigma^{2}, \tau) \left(\prod_{i=1}^{N} P(\bm{\beta}_{i} \lvert \bm{\mu}, \bm{\Sigma})\right) P(r \lvert r_{0},h) \dots\\
        & \dots P(h\lvert b_{0},c_{0})\left(\prod_{i=1}^N P(\omega_{i}\lvert r)\right)   P(\bm{\gamma} \lvert \bm{\zeta}_{\bm{\gamma}}, \bm{\Delta}_{\bm{\gamma}}) P(\sigma^{-2}  \lvert b_{\sigma^2},c_{\sigma^2})\dots \\
        & \dots P(\tau \lvert \zeta_{\tau}, \sigma_{\tau}^2) P(\bm{\mu} \lvert \bm{\zeta}_{\bm{\mu}}, \bm{\Delta}_{\bm{\mu}}) \left(\prod_{k=1}^K P(a_{k} \lvert s, \eta_{k})\right) P(\bm{\Sigma} \lvert \rho, \bm{B}).
    \end{split}
\end{equation}
Finally, to obtain conjugate posterior updates of the dispersion parameter $r$, we use a compound Poisson representation of negative binomial distribution (see Appendix \ref{app:r_mcmc}).    

\subsection{Markov chain Monte Carlo estimation} \label{sec:mcmc} 

MCMC estimation approximates a posterior distribution of interest through simulation of a Markov chain. In the present application, a Markov chain can be constructed by iteratively sampling from the conditional distributions of the parameters collected in $\bm{\Theta}$. As a results of P{\'o}lya-Gamma data augmentation, the conditional distributions of all model parameters, with the exception of the conditional distribution of the spatial association parameter $\tau$, are conjugate to their prior and belong to known families of standard parametric distribution. Since the conditional distribution of $\tau$ does not correspond to any recognisable distribution, we adopt the random-walk Metropolis algorithm to generate samples of it. The resulting Gibbs sampler is presented in Algorithm \ref{alg:gibbs}. In the algorithm, $\varpi_{\tau}$ is the step size of the random-walk Metropolis algorithm, which needs to be tuned.

\begin{algorithm}[!ht]
\begin{small}
\SetAlgoLined
\SetKwFor{For}{for}{sample from}{end for}%
\textbf{Initialization:}\\
Set hyper-parameters: $\{\bm{\zeta}_{\bm{\mu}}, \bm{\Delta}_{\bm{\mu}},\bm{\zeta}_{\bm{\gamma}}, \bm{\Delta}_{\bm{\gamma}}, \zeta_{\tau}, \sigma_{\tau}^2,  b_{\sigma^2},c_{\sigma^2}, r_{0}, b_{0},c_{0}, \nu, \{A_{k}\}_{k=1}^K \}$ \;
Initialize latent variables: $\left\{\bm{\phi}, \bm{\gamma}, \bm{\beta}, \bm{\mu}, \bm{a},\bm{\Sigma}, \sigma^{2}, \bm{\omega}, r, h, \tau \right\}$ \; 
        \For{1 \KwTo max-iteration}{        
          $\bm{\phi} \vert - \sim \text{Normal}\left ( (\bm{\Omega} + \Tilde{\bm{\Omega}})^{-1}\bm{\Omega}(\bm{Z} -\bm{M} \bm{\gamma}  - \bm{X} \bm{\beta}), (\bm{\Omega} + \Tilde{\bm{\Omega}})^{-1}\right)$  \; 
          $\bm{\gamma}  \vert - \sim \text{Normal}\left ((\bm{\Delta}_{\bm{\gamma}}^{-1} + \bm{M}^{\top}\bm{\Omega}\bm{M})^{-1}[\bm{M}^{\top} \bm{\Omega} (\bm{Z} - \bm{X} \bm{\beta} - \bm{\phi}) + \bm{\Delta}_{\bm{\gamma}}^{-1}\bm{\zeta}_{\bm{\gamma}}], (\bm{\Delta}_{\bm{\gamma}}^{-1} + \bm{M}^{\top}\bm{\Omega}\bm{M})^{-1}\right)$  \;
         $\{\bm{\beta}_i \lvert - \}_{i=1}^{N} \sim \text{Normal}\left(\left(\left[\omega_{i}\bm{X}_{i}\bm{X}_{i}^{\top}\right]^{-1} + \bm{\Sigma}\right) \left[ \omega_{i} (Z_{i} - \bm{M}_{i}^{\top} \bm{\gamma} -\phi_{i}) \bm{X}_{i}  + \bm{\Sigma}^{-1} \bm{\mu} \right],\left[\omega_{i}\bm{X}_{i}\bm{X}_{i}^{\top}\right]^{-1} + \bm{\Sigma}
        \right)$\; 
        $\bm{\mu}  \vert - \sim \text{Normal}\left((N \bm{\Sigma}^{-1} + \bm{\Delta}_{\bm{\mu}}^{-1})^{-1} \left( \bm{\Sigma}^{-1} \sum_{i=1}^{N} \bm{\beta}_{i}  + \bm{\Delta}_{\bm{\mu}}^{-1}\bm{\zeta}_{\bm{\mu}}  \right), 
       (N \bm{\Sigma}^{-1} + \bm{\Delta}_{\bm{\mu}}^{-1})^{-1} \right)$\; 
       $\{a_{k} \vert -\}_{k=1}^{K} \sim \text{Gamma} \left(\frac{\nu+K}{2}, \frac{1}{A_{k}^2} + \nu\left(\bm{\Sigma}^{-1}\right)_{kk} \right)$ \; 
       $\bm{\Sigma}  \vert - \sim \text{IW} \left( \nu + N + K -1, \bm{B} + \sum_{i=1}^N[\bm{\beta}_{i} -  \bm{\mu}][\bm{\beta}_{i} -  \bm{\mu}]^{\top}\right)$ \; 
       $\sigma^{-2}  \vert - \sim \text{Gamma}\left(b_{\sigma^2} + \frac{N}{2}, c_{\sigma^2} + \frac{ \bm{\phi}^{\top} \bm{S}^{\top} \bm{S} \bm{\phi}}{2} \right)$ \;
       $\{\omega_{i}\vert - \}_{i=1}^{N} \sim \text{PG}(y_{i} + r, \psi_i) $ \; 
       $r  \vert - \sim \text{Gamma}\left(r_{0} + \sum_{i=1}^{N} L_{i}, h + \sum_{i=1}^{N} \ln(1+\exp({\psi}_{i})) \right)$ (see details in Appendix \ref{app:r_mcmc}) \;
       $h  \vert - \sim \text{Gamma}(r_{0} + b_{0} , r + c_{0}) $ \; 
       
       $\tau \vert -$ (random-walk Metropolis step)
         \begin{itemize}
         \itemsep0em
            \item Propose $\tilde{\tau} = \tau + \sqrt{\varpi_{\tau}} \sigma_{\tau} \varsigma$, where $\varsigma \sim \text{Normal}(0,1)$;
            \item Compute $\xi = 
            \frac{P(\tilde{\tau} \lvert \zeta_{\tau}, \sigma_{\tau}^2)  P(\bm{\phi}  \lvert \tilde{\tau}, \sigma^2)}
            {P(\tau \lvert \zeta_{\tau}, \sigma_{\tau}^2)  P(\bm{\phi}  \lvert \tau, \sigma^2)}$;
            \item Draw $u \sim \text{Uniform}(0,1)$. If $\xi \leq u$, accept the proposal, else reject it. 
         \end{itemize}
		 }
\caption{Gibbs sampler for posterior inference in the spatial negative binomial model} \label{alg:gibbs}
\end{small}
\end{algorithm}

\subsection{Variational Bayes estimation} \label{sec:VB} 

In this section, we propose a variational Bayesian (VB) method to estimate the spatial negative binomial regression model. The goal of VB is to find a variational distribution $q(\bm{\Theta})$, which approximates the posterior distribution of interest, via minimisation of the probability distance between the variational distribution and the actual posterior distribution \citep{jordan1999introduction,blei2017variational}. The probability distance is conveniently measured by Kullback-Leibler (KL) divergence, which is defined as follows: 
\begin{equation} 
\begin{split} \label{eq:KL}
\text{KL} \left (q(\bm{\Theta}) \vert \vert P(\bm{\Theta} \vert \bm{y}) \right )
& = \int \ln \left ( \frac{q(\bm{\Theta})}{P(\bm{\Theta} \vert \bm{y})} \right ) q(\bm{\Theta}) d \bm{\Theta} \\
& = \mathbb{E}_{q} \left [ \ln q(\bm{\Theta}) \right ] - \mathbb{E}_{q} \left [ \ln P(\bm{\Theta} \vert \bm{y}) \right ]\\
& = \mathbb{E}_{q} \left [ \ln q(\bm{\Theta}) \right ] - \mathbb{E}_{q} \left [ \ln P(\bm{\Theta}, \bm{y}) \right ] + \ln P(\bm{y}).
\end{split}
\end{equation}
VB aims to minimise the KL divergence, which implies that
\begin{equation} 
q^{*}(\bm{\Theta})
= \operatorname*{arg\,min}_{q} \; \;\text{KL} \left (q(\bm{\Theta}) \vert \vert P(\bm{\Theta} \vert \bm{y}) \right ) .
\end{equation}

However, since $\ln P(\bm{y})$ has no closed form expression, the KL divergence is not analytically tractable. Recognising that $\mathbb{E}_{q} \left [ \ln q(\bm{\Theta}) \right ] - \mathbb{E}_{q} \left [ \ln P(\bm{\Theta}, \bm{y}) \right ]$ is negative of the evidence lower bound (ELBO), we rearrange Equation \ref{eq:KL} as follows:   
\begin{equation}\label{eq:elbo}
    \text{ELBO} = \ln P(\bm{y}) - \text{KL} \left (q(\bm{\Theta}) \vert \vert P(\bm{\Theta} \vert \bm{y}) \right ).
\end{equation}
Since the KL divergence is always positive, equation \ref{eq:elbo} shows that the optimal variational distribution can be equivalently obtained by maximising the ELBO.  

The variational distribution must be selected by the analyst. Its specification determines both the quality of the posterior approximation as well as the complexity of the optimisation problem \citep{blei2017variational}. In the following subsections, we describe two approaches for the specification of the variational distribution and suitable methods for ELBO maximisation.

\subsubsection{Mean field variational Bayes (MFVB)} 

MFVB specifies the density of the variational distribution as a product of the component-specific variational densities:  
\begin{equation}
    q(\bm{\Theta}) = \prod_{j=1}^{J} q(\bm{\Theta}_{j}),
\end{equation}
where $j \in \{1, \ldots, J\}$ are indexes of model parameter blocks. This specification imposes posterior independence between blocks of model parameters. The optimal variational density of a latent factor can be obtained using the following expression \citep{ormerod2010explaining}: 
\begin{equation} \label{eq:vbconj}
q^{*}(\bm{\Theta}_{j}) \propto \exp \left(\mathbb{E}_{- \bm{\Theta}_{j}} \left [ \ln P(\bm{y}, \bm{\Theta}) \right ]\right).
\end{equation}
If the conditional conjugacy holds for a model parameter, its variational distribution belongs to a recognisable family and can be easily obtained using the above equation. In case of non-conjugacy, the optimal variational density  $q^{*}(\bm{\Theta}_{j})$ of a model parameters can be obtained using quasi-Newton methods, non-conjugate variational message passing \citep{knowles2011non}, stochastic linear regression \citep{salimans2013fixed}, or Laplace approximation \citep[see][for a comprehensive review]{wang2013variational}. 

In the P{\'o}lya-Gamma-augmented spatial NB model, the conditional conjugacy holds for all model parameters, except for $\tau$. We thus obtain the optimal variational density of $\tau$ using non-conjugate variational message passing, while the optimal variational density of the remaining model parameters are obtained using equation \ref{eq:vbconj}. The results of MFVB indicate that the variational distributions of all variables, except $\tau$ and $\sigma^{2}$, closely resemble the posterior estimates of MCMC. This observation is well aligned with the findings of \cite{wu2018fast} in linear spatial models. However, in accordance with \cite{wu2018fast}, we also find that $\tau$ and $\sigma^{2}$ are poorly recovered by MFVB because of the untenable assumption of posterior independence. 

\subsubsection{Integrated non-factorised variational Bayes (INFVB)}

To address the bottlenecks of MFVB in the estimation of the considered spatial NB model, we propose INFVB method \citep{han2013integrated,wu2018fast}. INFVB decomposes latent variables $\bm{\Theta}$ into two disjoint subsets $\{\bm{\Theta}_{c},\bm{\Theta}_{d}\}$ to specify a flexible variational distribution:  
\begin{equation}
    q_{\text{INFVB}}(\bm{\Theta}) = q(\bm{\Theta}_{c} \lvert \bm{\Theta}_{d})q(\bm{\Theta}_{d}). 
\end{equation}
Since direct maximization of ELBO to find optimal variational density $q^{*}_{\text{INFVB}}(\bm{\Theta})$ is computationally challenging, a discrete distribution is specified on $\bm{\Theta}_{d}$ by discretising its domain using a multi-dimensional grid. We adopt a two-step procedure to obtain the optimal variational density $q^{*}_{\text{INFVB}}(\bm{\Theta})$: 
\begin{enumerate}
    \item For each grid point $\bm{\Theta}^{(g)}_{d} \in \{\bm{\Theta}^{(1)}_{d}, \dots,  \bm{\Theta}^{(G)}_{d}\}$, we obtain $q^{*}(\bm{\Theta}^{(g)}_{c} \lvert \bm{\Theta}^{(g)}_{d})$ and  $q^{*}(\bm{\Theta}^{(g)}_{d})$ (up to a multiplicative constant) using equations \ref{eq:invb_C} and \ref{eq:invb_D}, respectively \citep{han2013integrated}:
\begin{equation} \label{eq:invb_C}
     q^{*}\left(\bm{\Theta}^{(g)}_{c} \lvert \bm{\Theta}^{(g)}_{d}\right) =  \operatorname*{arg\,min}_{q\left(\bm{\Theta}^{(g)}_{c} \lvert \bm{\Theta}^{(g)}_{d}\right)}  \mathbb{E}_{q} \left [ \ln q\left(\bm{\Theta}^{(g)}_{c} \lvert \bm{\Theta}^{(g)}_{d}\right) \right ] - \mathbb{E}_{q} \left [ \ln P\left(\bm{y}, \bm{\Theta}^{(g)}_{c},\bm{\Theta}^{(g)}_{d}\right) \right ],
\end{equation}
\begin{equation}\label{eq:invb_D}
        q^{*}\left(\bm{\Theta}^{(g)}_{d}\right) \propto \exp\left(\mathbb{E} \left[ \ln P\left(\bm{y} , \bm{\Theta}^{(g)}_c,\bm{\Theta}^{(g)}_{d} \right) \right] - \mathbb{E}\left[\ln q^{*}\left(\bm{\Theta}^{(g)}_{c} \lvert \bm{\Theta}^{(g)}_{d}\right)\right]\right) .
\end{equation}
    \item We then compute optimal variational densities of $\bm{\Theta}_{d}$ and $\bm{\Theta}_{c}$ using equation \ref{eq:invb_CD}: 
    \begin{equation} \label{eq:invb_CD}
    \begin{split}
        & q^{*}(\bm{\Theta}_{d}) = \sum_{g=1}^{G} q^{*}\left(\bm{\Theta}^{(g)}_{d}\right) \mathbb{1}\left(\bm{\Theta}_{d} = \bm{\Theta}^{(g)}_{d}\right),  \\ 
        & q^{*}(\bm{\Theta}_{c}) = \sum_{g=1}^{G} q^{*}\left(\bm{\Theta}^{(g)}_{d}\right) q^{*}\left(\bm{\Theta}^{(g)}_{c} \lvert \bm{\Theta}^{(g)}_{d}\right), \\
        \text{where} \quad & q^{*}\left(\bm{\Theta}^{(g)}_{d}\right) = \frac{\exp\left(\mathbb{E} \left[ \ln P\left(\bm{y} , \bm{\Theta}^{(g)}_c,\bm{\Theta}^{(g)}_{d} \right) \right] - \mathbb{E}\left[\ln q^{*}\left(\bm{\Theta}^{(g)}_{c} \lvert \bm{\Theta}^{(g)}_{d}\right)\right]\right)}{\sum_{e=1}^{G} \exp\left(\mathbb{E} \left[ \ln P\left(\bm{y} , \bm{\Theta}^{(e)}_c,\bm{\Theta}^{(e)}_{d} \right) \right] - \mathbb{E}\left[\ln q^{*}\left(\bm{\Theta}^{(e)}_{c} \lvert \bm{\Theta}^{(e)}_{d}\right)\right]\right)}.
    \end{split}
    \end{equation}
\end{enumerate}

We highlight three important features of INFVB. First, the optimal density update of $\bm{\Theta}^{(g)}_{c} \lvert \bm{\Theta}^{(g)}_{d}$ using equation \ref{eq:invb_C} results into similar updates as obtained in MFVB (see equation \ref{eq:vbconj}). As a consequence, computation of $q^{*}\left(\bm{\Theta}^{(g)}_{c} \lvert \bm{\Theta}^{(g)}_{d}\right)$ is straightforward if conditional conjugacy holds for $\bm{\Theta}_{c}$. Second, the first step of INFVB includes embarrassingly parallel tasks. The communications overhead of these tasks is negligible, because the results of each task are only combined once during estimation. These characteristics make INFVB computationally efficient and scalable for large datasets. Third, if we consider $\bm{\Theta}_{d}$ as a vector of hyper-parameters, INFVB can be viewed as a generalised version of INLA. Specifically, INFVB relaxes the INLA's strict assumption on the normality of the conditional distribution $q(\bm{\Theta}_{c} \lvert \bm{\Theta}_{d})$ \citep[see section 2.3 of][for a detailed discussion on the superiority of INFVB over INLA]{han2013integrated}.

\subsubsection{INFVB for the spatial negative binomial model} 

On the basis of the findings of MFVB, we consider $\bm{\Theta}_{d} = \{\tau, \sigma^2\}$ and $\bm{\Theta}_{c} = \bm{\Theta} \setminus \bm{\Theta}_{d}$. We specify a nonparametric distribution on $\bm{\Theta}_{d}$ by discretising its domain using a two-dimensional grid and consider the following product form representation of $q(\bm{\Theta}_{c})$:  
\begin{equation}
\begin{split}
    q(\bm{\Theta}_{c}) & = q(\bm{\phi} \lvert \bm{\lambda}_{\bm{\phi}}, \bm{\Lambda}_{\bm{\phi}}) q(\bm{\gamma} \lvert  \bm{\lambda}_{\bm{\gamma}}, \bm{\Lambda}_{\bm{\gamma}}) q(\bm{\beta} \lvert  \bm{\lambda}_{\bm{\beta}}, \bm{\Lambda}_{\bm{\beta}}) q(\bm{\mu} \lvert  \bm{\lambda}_{\bm{\mu}}, \bm{\Lambda}_{\bm{\mu}}) \prod_{k=1}^{K}q({a}_{k} \lvert \Tilde{b}_{a_{k}}, \Tilde{c}_{a_{k}}) \dots \\
    & \dots q(\bm{\Sigma} \lvert  \Tilde{\rho}, \Tilde{\bm{B}}) \prod_{i=1}^N q(\omega_{i} \lvert \Tilde{b}_{\omega_{i}}, \Tilde{c}_{\omega_{i}}) q(h \lvert \Tilde{b}_{h}, \Tilde{c}_{h}) q(r) \prod_{i=1}^{N} q(L_{i}). 
\end{split}
\end{equation}
We find that variational distributions of model parameters blocks in $\bm{\Theta}_{c}$ belong to known families of distributions due to conjugacy: 
\begin{alignat*}{3}
&  q(\bm{\phi})  \sim \text{Normal}(\bm{\lambda}_{\bm{\phi}}, \bm{\Lambda}_{\bm{\phi}}), \quad \quad \quad &&  q(\bm{\gamma})\sim \text{Normal}( \bm{\lambda}_{\bm{\gamma}}, \bm{\Lambda}_{\bm{\gamma}}), \quad \quad \quad && \{ q(\bm{\beta}_{i})\}_{i=1}^{N} \sim \text{Normal}(\bm{\lambda}_{\bm{\beta}_{i}}, \bm{\Lambda}_{\bm{\beta}_{i}}),  \\
&  q(\bm{\mu})  \sim \text{Normal}(\bm{\lambda}_{\bm{\mu}}, \bm{\Lambda}_{\bm{\mu}}), \quad \quad \quad && \{ q(a_{k})\}_{k=1}^{K} \sim \text{Gamma}(\Tilde{b}_{a_{k}}, \Tilde{c}_{a_{k}}), \quad \quad \quad &&  q(\bm{\Sigma}) \sim \text{IW}(\Tilde{\rho}, \Tilde{\bm{B}}), \\
& \{ q(\omega_{i})\}_{i=1}^{N}   \sim \text{PG}(\Tilde{b}_{\omega_{i}}, \Tilde{c}_{\omega_{i}}), \quad \quad \quad &&  q(h) \sim \text{Gamma}( \Tilde{b}_{h}, \Tilde{c}_{h}), \quad \quad \quad &&  q(r) \sim \text{Gamma}( \Tilde{b}_{r}, \Tilde{c}_{r}), \\
& \{q(L_{i})\}_{i=1}^{N} = \sum_{j=0}^{y_{i}} R_{\tilde{r}}(y_{i},j) \delta_{j}, \quad \quad \quad &&  q(\bm{\psi}) \sim \text{Normal}(\bm{\lambda}_{\bm{\psi}}, \bm{\Lambda}_{\bm{\psi}}).
\end{alignat*}

\begin{algorithm}[!ht]
\begin{small}
\SetAlgoLined
\SetKwInOut{Input}{Step}\SetKwInOut{Output}{Step}
\SetKwFor{For}{for}{obtain $q^{*}\left(\bm{\Theta}^{(g)}_{c} \lvert \bm{\Theta}^{(g)}_{d}\right)$ and  $q^{*}\left(\bm{\Theta}^{(g)}_{d}\right)$ in parallel}{end for}%
Set hyper-parameters: $\{\bm{\zeta}_{\bm{\mu}}, \bm{\Delta}_{\bm{\mu}},\bm{\zeta}_{\bm{\gamma}}, \bm{\Delta}_{\bm{\gamma}}, \zeta_{\tau}, \sigma_{\tau}^2,  b_{\sigma^2},c_{\sigma^2}, r_{0}, b_{0},c_{0}, \nu, \{A_{k}\}_{k=1}^K \}$ \;
Compute fixed variational parameters: $\Tilde{b}_{a_k} = \frac{\nu+K}{2}; \quad$  $\Tilde{\rho} =  \nu + N + K -1; \quad$  $\Tilde{b}_{h}  = r_{0} + b_{0}$ \; 
Specify a two-dimensional grid $\bm{\Theta}^{(g)}_{d} \in \{\bm{\Theta}^{(1)}_{d}, \dots,  \bm{\Theta}^{(G)}_{d}\}$ on the domain of $\bm{\Theta}_{d} = \{\tau, \sigma^2\}$ \; 
\BlankLine
\BlankLine
\Input{1}
\For{$g$ in 1 \KwTo $G$}{
\BlankLine
Initialize $\left\{\bm{\lambda}_{\bm{\phi}}^{(g)}, \bm{\Lambda}_{\bm{\phi}}^{(g)},\bm{\lambda}_{\bm{\gamma}}^{(g)}, \bm{\Lambda}_{\bm{\gamma}}^{(g)}, \left\{\bm{\lambda}_{\bm{\beta}_{i}}^{(g)}, \bm{\Lambda}_{\bm{\beta}_{i}}^{(g)}\right\}_{i=1}^N, \bm{\lambda}_{\bm{\mu}}^{(g)}, \bm{\Lambda}_{\bm{\mu}}^{(g)},\left\{\Tilde{c}_{a_{k}}^{(g)}\right\}_{k=1}^{K}, \Tilde{\bm{B}}^{(g)},\Tilde{c}_{h}^{(g)}, \Tilde{b}_{r}^{(g)}, \Tilde{c}_{r}^{(g)}\right\}$\;  
\While{not converged}{
$\bm{\Lambda}_{\bm{\phi}}^{(g)} = \left(\mathbb{E}[\bm{\Omega}]^{(g)} + \Tilde{\bm{\Omega}}^{(g)}\right)^{-1}$\;
$\bm{\lambda}_{\bm{\phi}}^{(g)} = \bm{\Lambda}_{\bm{\phi}}^{(g)} \left(\mathbb{E}\left[\{\bm{Z}^{*}\}^{(g)}\right] -\mathbb{E}[\bm{\Omega}^{(g)}]\bm{M} \bm{\lambda}_{\bm{\gamma}}^{(g)} - \mathbb{E}[\bm{\Omega}^{(g)}]\bm{X} \bm{\lambda}_{\bm{\beta}}^{(g)}\right)$\;
$\bm{\Lambda}_{\bm{\gamma}}^{(g)} = \left(\bm{\Delta}_{\bm{\gamma}}^{-1} + \bm{M}^{\top}\mathbb{E}\left[\bm{\Omega}^{(g)}\right]\bm{M}\right)^{-1} $\;
$\bm{\lambda}_{\bm{\gamma}}^{(g)} = \bm{\Lambda}_{\bm{\gamma}}^{(g)}\left(\bm{M}^{\top}  \left(\mathbb{E}[\{\bm{Z}^{*}\}^{(g)}] - \mathbb{E}\left[\bm{\Omega}^{(g)}\right]\bm{X} \bm{\lambda}_{\bm{\beta}}^{(g)} - \mathbb{E}\left[\bm{\Omega}^{(g)}\right]\bm{\lambda}_{\bm{\phi}}^{(g)}\right) + \bm{\Delta}_{\bm{\gamma}}^{-1}\bm{\zeta}_{\bm{\gamma}}\right)$ \;
$\left\{\bm{\Lambda}_{\bm{\beta}_{i}}^{(g)}\right\}_{i=1}^{N} = \left(\mathbb{E}\left[\omega_{i}^{(g)}\right]\bm{X}_{i}\bm{X}_{i}^{\top} + \tilde{\rho}\{\bm{\tilde{B}}^{(g)}\}^{-1}\right)^{-1}$ \; 
$\left\{\bm{\lambda}_{\bm{\beta}_{i}}^{(g)}\right\}_{i=1}^{N} = \bm{\Lambda}_{\bm{\beta}_{i}}^{(g)}\left[\left(\mathbb{E}\left[\{Z_{i}^{*}\}^{(g)}\right] - \mathbb{E}\left[\omega_{i}^{(g)}\right] \bm{M}_{i}^{\top} \bm{\lambda}_{\bm{\gamma}}^{(g)} - \mathbb{E}\left[\omega_{i}^{(g)}\right] \bm{\lambda}_{\phi_{i}}^{(g)}\right) \bm{X}_{i} +  \tilde{\rho}\{\bm{\tilde{B}}^{-1} \bm{\lambda}_{\bm{\mu}}\}^{(g)} \right]$\;
$\bm{\Lambda}_{\bm{\mu}}^{(g)} = \left[ N\tilde{\rho}\{\bm{\tilde{B}}^{(g)}\}^{-1} + \bm{\Delta}_{\bm{\mu}}^{-1}\right]^{-1}$ \;
$\bm{\lambda}_{\bm{\mu}}^{(g)} = \bm{\Lambda}_{\bm{\mu}}^{(g)}\left[\left(\tilde{\rho}\{\bm{\tilde{B}}^{(g)}\}^{-1}\right) \sum_{i=1}^N \bm{\lambda}_{\bm{\beta}_{i}}^{(g)}  + \bm{\Delta}_{\bm{\mu}}^{-1}\bm{\zeta}_{\bm{\mu}}  \right]$\;
$\left\{\Tilde{c}_{a_k}^{(g)}\right\}_{k=1}^{K} = \left[\frac{1}{A_{k}^2} + \nu  \tilde{\rho} \left(\{\bm{\tilde{B}}^{(g)}\}^{-1}\right)_{kk} \right]$ \;
$\Tilde{\bm{B}}^{(g)} = 2\nu \text{diag}\left( \frac{\tilde{b}_{\bm{a}}}{\tilde{c}_{\bm{a}}^{(g)}} \right) + N \bm{\Lambda}_{\bm{\mu}}^{(g)} +  \sum_{i=1}^N \left( \bm{\Lambda}_{\bm{\beta}_{i}} + [\bm{\lambda}_{\bm{\beta}_{i}} -  \bm{\lambda}_{\bm{\mu}}][\bm{\lambda}_{\bm{\beta}_{i}} -  \bm{\lambda}_{\bm{\mu}}]^{\top} \right)^{(g)}$\;
$\Tilde{c}_{h}^{(g)}  = \left(\frac{\tilde{b}_{r}}{\tilde{c}_{r}}\right)^{(g)} + c_{0}$\; 
$\Tilde{b}_{r}^{(g)} = r_{0} + \sum_{i=1}^{N} \mathbb{E}(L_{i}^{(g)})$ \;
$\Tilde{c}_{r}^{(g)} = \frac{\Tilde{b}_{h}}{\Tilde{c}_{h}^{(g)}} + \sum_{i=1}^{N} \mathbb{E}\left[\log\left(1+\exp\left(\psi_{i}^{(g)}\right)\right)\right]$ \;
$\bm{\lambda}_{\bm{\psi}}^{(g)} = \bm{M} \bm{\lambda}_{\bm{\gamma}}^{(g)} + \bm{X} \bm{\lambda}_{\bm{\beta}}^{(g)} + \bm{\lambda}_{\bm{\phi}}^{(g)} $ \;
$\bm{\Lambda}_{\bm{\psi}}^{(g)} =\bm{M} \bm{\Lambda}_{\bm{\gamma}}^{(g)} \bm{M}^{\top}  + \bm{X} \bm{\Lambda}_{\bm{\beta}}^{(g)}   \bm{X}^{\top}+  \bm{\Lambda}_{\bm{\phi}}^{(g)}$\;
}
Compute $q^{*}\left(\bm{\Theta}^{(g)}_{d}\right)$ up to a multiplicative constant by inserting expectations computed using equation \ref{eq:theta_d_upd} (see appendix \ref{app:q_d}) into equation \ref{eq:invb_D}\; 
}
\BlankLine
\BlankLine
\Input{2}
Obtain optimal variational densities of $\bm{\Theta}_{d}$ and $\bm{\Theta}_{c}$ using equation \ref{eq:invb_CD}\;
\caption{Integrated non-factorized variational Bayes (INFVB) method for the spatial NB model} \label{alg:infvb}
\end{small}
\end{algorithm}

We reiterate that a compound Poisson representation of negative binomial distribution is used to ensure conjugate posterior updates for the dispersion parameter $r$ (see Appendix \ref{app:r_mcmc} for details). Accordingly, we adopt the variational distribution used by \cite{zhou2012lognormal} on $L_{i}$, where $\delta_{j}$ is an indicator. 
The INFVB method to estimate the spatial count model is summarised in Algorithm \ref{alg:infvb}; supplementary identities and expressions are presented in Appendix \ref{app:imp_id}. The expression for the conditional ELBO, i.e. the negative of the function minimised in equation \ref{eq:invb_C} is presented in Appendix \ref{app:q_d}.    

\section{Simulation study} \label{sec:sim}

To evaluate computational efficiency and finite sample properties of INFVB and MCMC estimators, we conduct a Monte Carlo study. In this section, we present details of the data generating process (DGP), followed by performance measures, implementation details and results of the simulation study. 

\subsection{Data and experimental setup}

We generate data according to the following DGP:
\begin{align*}
& \bm{\beta}_{i}  \sim \text{Normal}(\bm{\mu}, \bm{\Sigma}), & & i = 1,\dots,N \\
& \bm{\epsilon} \sim \text{Normal}(0,\sigma^2 \bm{I}_{N}) ], \\
& \bm{S}\bm{\phi} = \exp(\tau \bm{W})\bm{\phi} =  \bm{\epsilon}, \\
& \psi_i = \bm{M}_{i}^{\top} \bm{\gamma} + \bm{X}_{i}^{\top} \bm{\beta}_{i} + \phi_{i},  & & i = 1,\dots, N \\
& p_i = \frac{\exp(\psi_i)}{1+\exp(\psi_i)}, & & i = 1,\dots,N \\
& y_{i} \sim \text{NB}(r, p_i). & & i = 1,\dots,N
\end{align*} 
We consider eight simulation scenarios defined through combinations of $N = \{1000, 1500\}$, $\tau = \{-0.7,0.7\}$, and $\sigma = \{0.2,0.4\}$. Ten resamples of each simulation scenario are generated, i.e. we estimate the spatial NB model using MCMC and INFVB on a total of 80 simulated datasets. For all simulation scenarios, we set 
$\bm{\mu} = \begin{bmatrix} 0.2 & -0.2 & 0.2 \end{bmatrix}^{\top}$, $\bm{\Sigma} = \text{diag}(\tilde{\bm{\sigma}}) \tilde{\bm{\Omega}} \text{diag}(\tilde{\bm{\sigma}})$ with $\tilde{\bm{\sigma}} = \begin{bmatrix} 0.141 & 0.141 & 0.141 \end{bmatrix}^{\top}$ and $\tilde{\bm{\Omega}} = \begin{bmatrix} 1 &  0.2 & 0 \\ 0.2 & 1 & 0.2 \\ 0 & 0.2 & 1 \end{bmatrix}^{\top}$ as well as $\bm{\gamma} = \begin{bmatrix} 1.0 & 0.3 & -0.3 & 0.3 \end{bmatrix}^{\top}$, and $r = 1.5$. Furthermore, we let $M_{i,1} = 1$ and $M_{i,q} \sim \text{Normal}(0,1)$ for $q = 2, 3, 4$ as well as $X_{i,k} \sim \text{Normal}(0,1)$ for $k = 1, 2, 3$. To construct the row-normalised spatial weights matrix $\bm{W}$, we calculate an 8-nearest neighbour matrix for $N$ points, which are randomly located in a unit square.

\subsection{Performance metrics} 

We evaluate the estimation accuracy of the INFVB and MCMC methods by calculating the mean of the absolute percent bias (APB) of model parameters across resamples. APB is a normalised measure of the finite sample bias and is given by $\text{APB} = \left \lvert \frac{\text{MPM} - \text{True value} }{\text{True value}} \right \lvert \times 100 $, where the mean posterior mean (MPM) is the average of the posterior mean across resamples. In addition, we also report the standard deviation of the posterior mean (SDPM) and the mean of posterior standard deviation (MPSD) across resamples.

\subsection{Implementation and estimation practicalities} \label{sec:imp_est_prac}

We implement the MCMC and INFVB methods for the spatial NB model by writing our own Python code. To draw from the P{\'o}lya-Gamma distribution, we use an existing implementation \citep{linderman2015dependent, linderman2016bayesian, linderman2016recurrent} of the sampling techniques proposed by \cite{polson2013bayesian} and \cite{windle2014sampling}.\footnote{The estimation code is publicly available at \url{https://github.com/RicoKrueger/infvb_spatial_count}.}

The MCMC sampler is executed with two parallel Markov chains and 40,000 iterations for each chain, whereby the initial 20,000 iterations are discarded for burn-in. After burn-in, every fifth draw is retained. The random-walk Metropolis step to generate samples from the conditional distribution of the spatial association parameter $\tau$ is adaptively scaled such that the average acceptance rate is approximately 44\%, which is the recommended acceptance ratio for a uni-dimensional target density \citep[see][]{roberts1997weak}. Convergence of the MCMC simulation is assessed with the help of the potential scale reduction factor \citep{gelman1992inference}.

For INFVB, a two-dimensional search space over $\{ \tau, \sigma \}$ is defined via the Cartesian product of two uni-dimensional grids. The grid over $\tau$ consists of 15 equidistant points in the interval $[0,1.4]$ or $[-1.4,0]$ (depending on the true value of $\tau$), while the grid over $\sigma$ consists of 10 equidistant points in the interval $[0.05, 0.8]$. We exploit the embarrassingly parallel computations of the INFVB method by distributing step 1 of Algorithm \ref{alg:infvb} over an eight-core processor.

\subsection{Results}

\begin{figure}[H]
\centering
\includegraphics[width = 0.7 \textwidth]{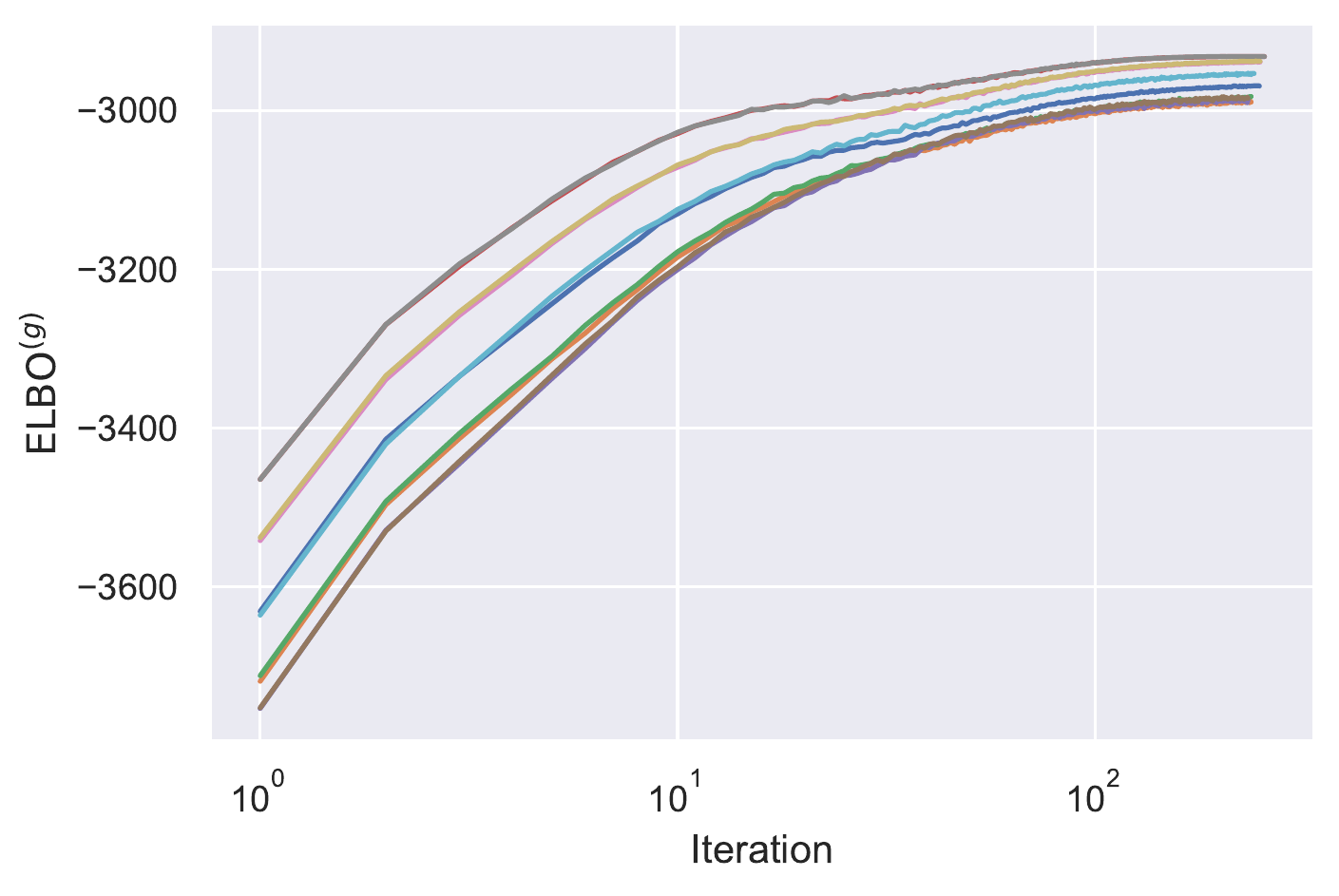}
\caption{Sequence of conditional ELBOs of ten randomly selected grid points for simulation scenario $\tau = -0.7$, $\sigma = 0.2$, $N = 1500$} \label{fig:elbo}
\end{figure}

Before comparing INFVB with MCMC, we demonstrate the accuracy of our analytical derivation and implementation of the INFVB method. In one resample of one specific simulation scenario, we plot the evolution of the conditional ELBO (presented in Appendix \ref{app:q_d}) over the number of iterations for ten randomly selected grid points in Figure \ref{fig:elbo}. It can be seen that the conditional ELBOs of the ten randomly grid points are monotonically increasing over iterations, which illustrates the correctness of the proposed INFVB estimator.

Table \ref{tab:estimation_time_synth} enumerates the computation times of the MCMC and INFVB estimators for all DGP instances. INFVB is approximately 50 times faster than MCMC for all instances of the DGP. Considerably low standard deviations of the estimation time across resamples underscore the robustness of this result. Further reductions in the estimation time of INFVB could be realised by distributing step 1 of Algorithm \ref{alg:infvb} over more than eight compute cores.  

Next, we present the results of the other performance measures for four out of the eight simulation scenarios in Tables \ref{tab:results_synth_N1500_S1} to \ref{tab:results_synth_N1500_S4}.\footnote{The results for the remaining for simulation scenarios with $N=1000$ offer similar insights and are thus included as supplementary material.} Similar and considerably low APB values (below 10\% for most of the parameters), and small SDPM values indicate that INFVB and MCMC not only recover the true parameters quite well but also with an identical precision across all the considered simulation scenarios. As an exception, the recovery of $\sigma$ is poor in INFVB and a similar bias is observed for $\tau$ in MCMC. However, both $\tau$ and $\sigma$ are recovered equally well by MCMC and INFVB in the empirical study (see Figure \ref{fig:posterior_real} in the next section). Furthermore, for most model parameters, MPSD is substantially lower for INFVB than for MCMC. This result corroborates the findings of earlier studies, which suggest that VB underestimates the posterior uncertainty \citep{blei2017variational,giordano2018covariances}.  

\begin{table}[H]
\centering
\small
\begin{tabular}{l|rr|rr}
\toprule
{} & \multicolumn{2}{c|}{INFVB} & \multicolumn{2}{c}{MCMC} \\
{} &  Mean & Std. dev. &   Mean & Std. dev. \\
\midrule
$N = 1000$ & & & & \\
\quad $\tau = -0.7$; $\sigma = 0.2$ &   9.1 &       0.2 &  494.0 &      17.6 \\
\quad $\tau = 0.7$; $\sigma = 0.2$  &   9.2 &       0.2 &  512.8 &       1.2 \\
\quad $\tau = -0.7$; $\sigma = 0.4$ &   9.4 &       0.1 &  525.3 &       4.1 \\
\quad $\tau = 0.7$; $\sigma = 0.4$  &   9.3 &       0.1 &  506.7 &       1.6 \\
\midrule
$N = 1500$ & & & & \\
\quad $\tau = -0.7$; $\sigma = 0.2$ &  28.2 &       0.3 & 1397.1 &      14.9 \\
\quad $\tau = 0.7$; $\sigma = 0.2$  &  28.2 &       0.4 & 1423.2 &       6.4 \\
\quad $\tau = -0.7$; $\sigma = 0.4$ &  29.0 &       0.4 & 1491.8 &      10.4 \\
\quad $\tau = 0.7$; $\sigma = 0.4$  &  21.5 &       1.1 & 1343.3 &       8.5 \\
\bottomrule
\end{tabular}

\caption{Estimation time in minutes across ten resamples by estimation method and simulation scenario} \label{tab:estimation_time_synth}
\end{table}

\begin{table}[H]
\centering
\small
\begin{tabular}{l|r|rrrr|rrrr}
\toprule
{} &  & \multicolumn{4}{c|}{INFVB} & \multicolumn{4}{c}{MCMC} \\
{} &   True &    MPM &  SDPM &  APB &  MPSD &    MPM &  SDPM &  APB &  MPSD \\
\midrule
$\gamma_{1}$         &  1.000 &  1.005 & 0.038 &  0.5 & 0.030 &  1.021 & 0.037 &  2.1 & 0.043 \\
$\gamma_{2}$         &  0.300 &  0.285 & 0.040 &  4.9 & 0.030 &  0.291 & 0.037 &  2.9 & 0.043 \\
$\gamma_{3}$         & -0.300 & -0.294 & 0.028 &  1.9 & 0.031 & -0.298 & 0.031 &  0.8 & 0.043 \\
$\gamma_{4}$         &  0.300 &  0.301 & 0.038 &  0.4 & 0.030 &  0.308 & 0.043 &  2.7 & 0.043 \\
$\mu_{1}$            &  0.200 &  0.197 & 0.021 &  1.3 & 0.003 &  0.202 & 0.022 &  1.2 & 0.026 \\
$\mu_{2}$            & -0.200 & -0.205 & 0.037 &  2.3 & 0.003 & -0.208 & 0.036 &  3.9 & 0.026 \\
$\mu_{3}$            &  0.200 &  0.199 & 0.034 &  0.4 & 0.003 &  0.205 & 0.037 &  2.6 & 0.026 \\
$\tilde{\sigma}_{1}$ &  0.141 &  0.123 & 0.017 & 13.0 & 0.004 &  0.146 & 0.064 &  3.2 & 0.057 \\
$\tilde{\sigma}_{2}$ &  0.141 &  0.120 & 0.013 & 15.4 & 0.004 &  0.135 & 0.053 &  4.3 & 0.065 \\
$\tilde{\sigma}_{3}$ &  0.141 &  0.116 & 0.011 & 18.1 & 0.004 &  0.111 & 0.044 & 21.7 & 0.061 \\
$\tau$               & -0.700 & -0.604 & 0.110 & 13.7 & 0.390 & -0.159 & 0.145 & 77.3 & 0.435 \\
$\sigma$             &  0.200 &  0.119 & 0.020 & 40.3 & 0.046 &  0.152 & 0.069 & 23.8 & 0.071 \\
$r$                  &  1.500 &  1.514 & 0.053 &  0.9 & 0.040 &  1.477 & 0.057 &  1.5 & 0.083 \\
\midrule 
\multicolumn{10}{p{12cm}}{ \footnotesize
Note:
MPM = mean of posterior mean;
SDPM = standard deviation of posterior mean;
APB = absolute percent bias;
MPSD = mean of posterior standard deviation.
All statistics are calculated across ten resamples.
} \\
\bottomrule
\end{tabular}

\caption{Simulation results for $\tau = -0.7$, $\sigma = 0.2$, $N = 1500$} \label{tab:results_synth_N1500_S1}
\end{table}

\begin{table}[H]
\centering
\small
\begin{tabular}{l|r|rrrr|rrrr}
\toprule
{} &  & \multicolumn{4}{c|}{INFVB} & \multicolumn{4}{c}{MCMC} \\
{} &   True &    MPM &  SDPM &  APB &  MPSD &    MPM &  SDPM &  APB &  MPSD \\
\midrule
$\gamma_{1}$         &  1.000 &  0.986 & 0.026 &  1.4 & 0.030 &  1.003 & 0.029 &   0.3 & 0.044 \\
$\gamma_{2}$         &  0.300 &  0.297 & 0.046 &  0.9 & 0.030 &  0.304 & 0.048 &   1.3 & 0.043 \\
$\gamma_{3}$         & -0.300 & -0.282 & 0.031 &  6.0 & 0.030 & -0.287 & 0.028 &   4.4 & 0.043 \\
$\gamma_{4}$         &  0.300 &  0.279 & 0.033 &  7.1 & 0.030 &  0.283 & 0.037 &   5.7 & 0.043 \\
$\mu_{1}$            &  0.200 &  0.184 & 0.023 &  7.8 & 0.003 &  0.192 & 0.023 &   3.8 & 0.027 \\
$\mu_{2}$            & -0.200 & -0.198 & 0.030 &  0.8 & 0.003 & -0.202 & 0.030 &   1.2 & 0.027 \\
$\mu_{3}$            &  0.200 &  0.202 & 0.027 &  0.8 & 0.003 &  0.204 & 0.030 &   2.2 & 0.027 \\
$\tilde{\sigma}_{1}$ &  0.141 &  0.122 & 0.013 & 13.8 & 0.004 &  0.134 & 0.056 &   5.6 & 0.065 \\
$\tilde{\sigma}_{2}$ &  0.141 &  0.129 & 0.016 &  8.7 & 0.004 &  0.150 & 0.057 &   6.2 & 0.070 \\
$\tilde{\sigma}_{3}$ &  0.141 &  0.118 & 0.014 & 16.5 & 0.004 &  0.132 & 0.042 &   6.9 & 0.058 \\
$\tau$               &  0.700 &  0.633 & 0.041 &  9.6 & 0.421 & -0.045 & 0.150 & 106.5 & 0.435 \\
$\sigma$             &  0.200 &  0.116 & 0.017 & 41.8 & 0.046 &  0.153 & 0.052 &  23.4 & 0.084 \\
$r$                  &  1.500 &  1.531 & 0.061 &  2.0 & 0.039 &  1.497 & 0.061 &   0.2 & 0.089 \\
\midrule
\multicolumn{10}{p{12cm}}{ \footnotesize
Note: For an explanation of the table headers see Table \ref{tab:results_synth_N1500_S1}.
} \\
\bottomrule
\end{tabular}

\caption{Simulation results for $\tau = 0.7$, $\sigma = 0.2$, $N = 1500$} \label{tab:results_synth_N1500_S2}
\end{table}

\begin{table}[H]
\centering
\small
\begin{tabular}{l|r|rrrr|rrrr}
\toprule
{} &  & \multicolumn{4}{c|}{INFVB} & \multicolumn{4}{c}{MCMC} \\
{} &   True &    MPM &  SDPM &  APB &  MPSD &    MPM &  SDPM &  APB &  MPSD \\
\midrule
$\gamma_{1}$         &  1.000 &  0.979 & 0.032 &  2.1 & 0.031 &  0.981 & 0.032 &  1.9 & 0.048 \\
$\gamma_{2}$         &  0.300 &  0.317 & 0.049 &  5.6 & 0.031 &  0.304 & 0.050 &  1.3 & 0.047 \\
$\gamma_{3}$         & -0.300 & -0.275 & 0.033 &  8.3 & 0.032 & -0.299 & 0.036 &  0.4 & 0.047 \\
$\gamma_{4}$         &  0.300 &  0.299 & 0.041 &  0.2 & 0.031 &  0.290 & 0.043 &  3.5 & 0.047 \\
$\mu_{1}$            &  0.200 &  0.199 & 0.026 &  0.7 & 0.004 &  0.206 & 0.028 &  2.9 & 0.029 \\
$\mu_{2}$            & -0.200 & -0.190 & 0.036 &  5.0 & 0.004 & -0.196 & 0.036 &  1.9 & 0.028 \\
$\mu_{3}$            &  0.200 &  0.203 & 0.030 &  1.5 & 0.004 &  0.208 & 0.032 &  4.0 & 0.029 \\
$\tilde{\sigma}_{1}$ &  0.141 &  0.127 & 0.017 & 10.5 & 0.008 &  0.152 & 0.072 &  7.4 & 0.066 \\
$\tilde{\sigma}_{2}$ &  0.141 &  0.126 & 0.016 & 10.9 & 0.008 &  0.135 & 0.044 &  4.4 & 0.071 \\
$\tilde{\sigma}_{3}$ &  0.141 &  0.126 & 0.018 & 10.9 & 0.007 &  0.152 & 0.054 &  7.5 & 0.070 \\
$\tau$               & -0.700 & -1.025 & 0.203 & 46.4 & 0.293 & -0.635 & 0.194 &  9.2 & 0.250 \\
$\sigma$             &  0.400 &  0.184 & 0.033 & 54.0 & 0.048 &  0.359 & 0.075 & 10.3 & 0.073 \\
$r$                  &  1.500 &  1.480 & 0.114 &  1.3 & 0.056 &  1.519 & 0.118 &  1.3 & 0.101 \\
\midrule
\multicolumn{10}{p{12cm}}{ \footnotesize
Note: For an explanation of the table headers see Table \ref{tab:results_synth_N1500_S1}.
} \\
\bottomrule
\end{tabular}

\caption{Simulation results for $\tau = -0.7$, $\sigma = 0.4$, $N = 1500$} \label{tab:results_synth_N1500_S3}
\end{table}

\begin{table}[H]
\centering
\small
\begin{tabular}{l|r|rrrr|rrrr}
\toprule
{} &  & \multicolumn{4}{c|}{INFVB} & \multicolumn{4}{c}{MCMC} \\
{} &   True &    MPM &  SDPM &  APB &  MPSD &    MPM &  SDPM &  APB &  MPSD \\
\midrule
$\gamma_{1}$         &  1.000 &  1.024 & 0.061 &  2.4 & 0.032 &  1.015 & 0.060 &  1.5 & 0.048 \\
$\gamma_{2}$         &  0.300 &  0.302 & 0.058 &  0.8 & 0.032 &  0.280 & 0.053 &  6.6 & 0.048 \\
$\gamma_{3}$         & -0.300 & -0.254 & 0.045 & 15.4 & 0.031 & -0.283 & 0.053 &  5.7 & 0.048 \\
$\gamma_{4}$         &  0.300 &  0.313 & 0.026 &  4.3 & 0.032 &  0.292 & 0.031 &  2.7 & 0.048 \\
$\mu_{1}$            &  0.200 &  0.193 & 0.028 &  3.3 & 0.004 &  0.203 & 0.037 &  1.7 & 0.031 \\
$\mu_{2}$            & -0.200 & -0.189 & 0.026 &  5.6 & 0.003 & -0.193 & 0.028 &  3.5 & 0.029 \\
$\mu_{3}$            &  0.200 &  0.205 & 0.027 &  2.4 & 0.003 &  0.211 & 0.030 &  5.6 & 0.028 \\
$\tilde{\sigma}_{1}$ &  0.141 &  0.133 & 0.020 &  5.8 & 0.008 &  0.160 & 0.066 & 13.3 & 0.073 \\
$\tilde{\sigma}_{2}$ &  0.141 &  0.128 & 0.019 &  9.3 & 0.007 &  0.134 & 0.052 &  5.0 & 0.065 \\
$\tilde{\sigma}_{3}$ &  0.141 &  0.123 & 0.016 & 13.1 & 0.007 &  0.128 & 0.051 &  9.5 & 0.068 \\
$\tau$               &  0.700 &  0.717 & 0.079 &  2.4 & 0.419 &  0.295 & 0.166 & 57.8 & 0.325 \\
$\sigma$             &  0.400 &  0.163 & 0.021 & 59.3 & 0.056 &  0.366 & 0.086 &  8.5 & 0.087 \\
$r$                  &  1.500 &  1.404 & 0.089 &  6.4 & 0.052 &  1.482 & 0.101 &  1.2 & 0.102 \\
\midrule
\multicolumn{10}{p{12cm}}{ \footnotesize
Note: For an explanation of the table headers see Table \ref{tab:results_synth_N1500_S1}.
} \\
\bottomrule
\end{tabular}

\caption{Simulation results for $\tau = 0.7$, $\sigma = 0.4$, $N = 1500$} \label{tab:results_synth_N1500_S4}
\end{table}

\section{Case study}\label{sec:app}
In this section, we compare the performance of INFVB and MCMC in terms of computational efficiency, goodness-of-fit, and marginal posterior distributions of model parameters in an empirical application. 

\subsection{Data}

\begin{figure}[H]
\centering
\includegraphics[width = 0.8 \textwidth]{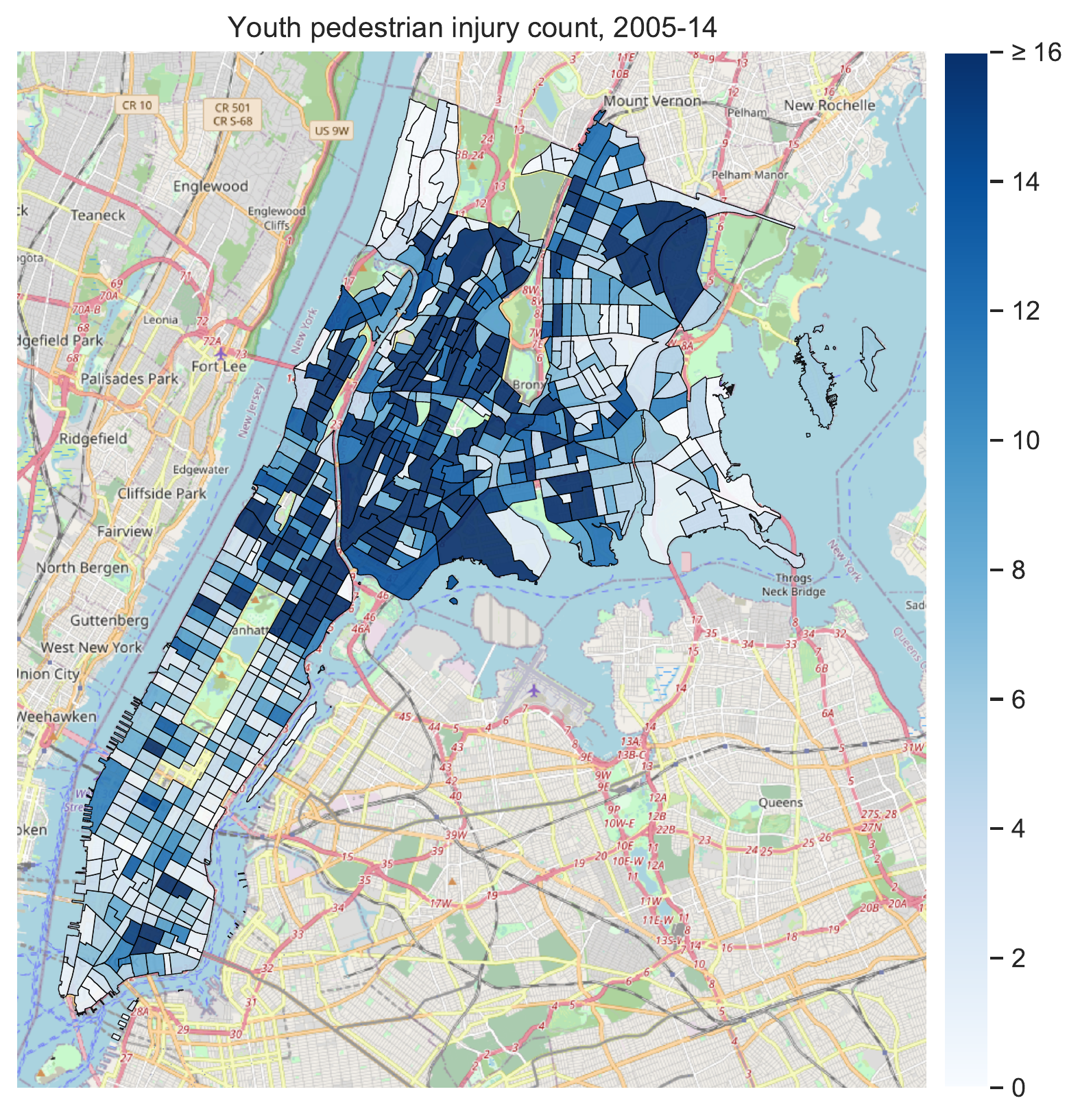}
\caption{Observed youth pedestrian injury counts in the Bronx and Manhattan in 2005-14 by census tract} \label{fig:map_obs_counts}
\end{figure}

\begin{figure}[H]
\centering
\includegraphics[width = 0.7 \textwidth]{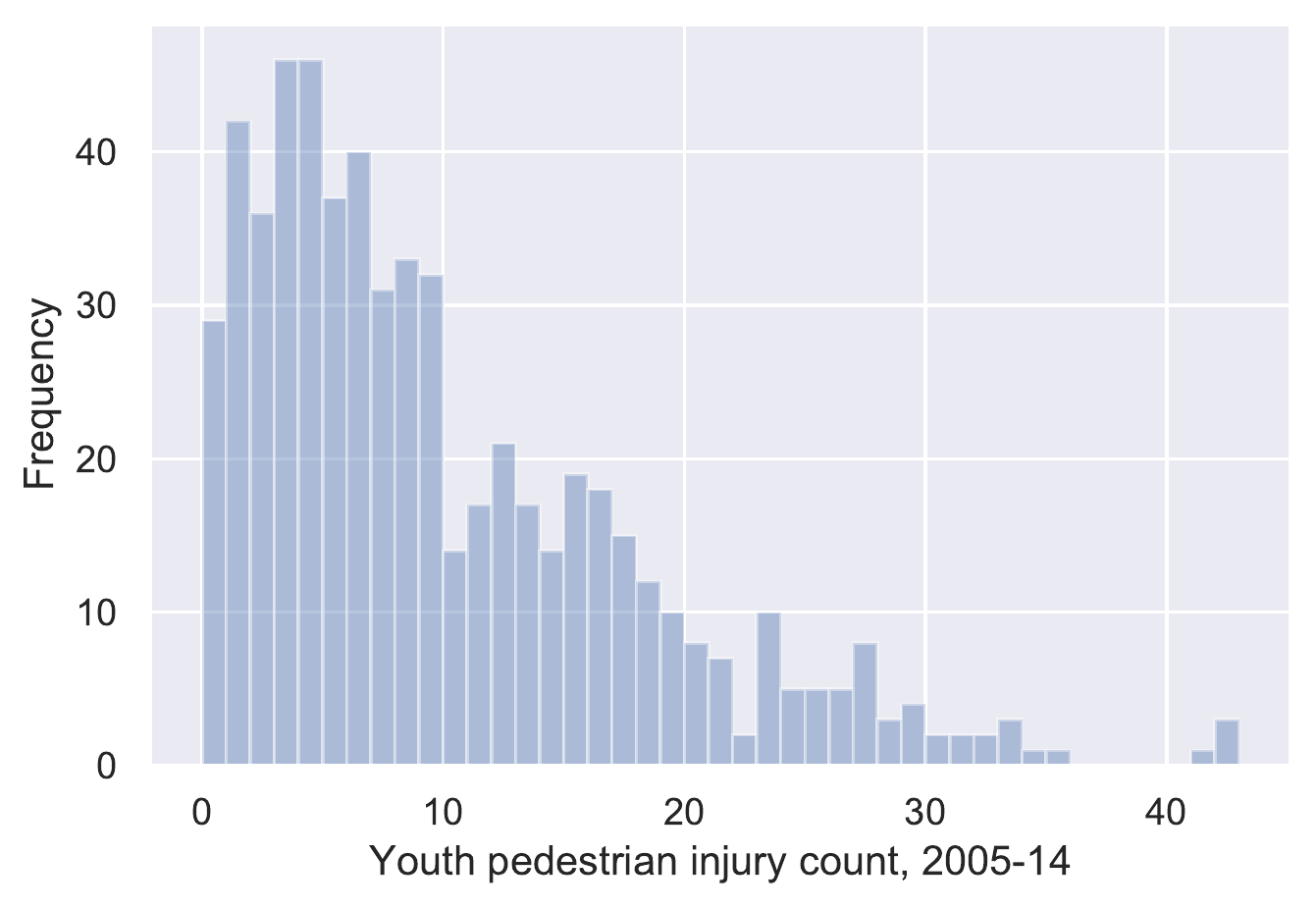}
\caption{Histogram of observed youth pedestrian injury counts in the Bronx and Manhattan in 2005-14 by census tract} \label{fig:hist_obs_counts}
\end{figure}

The data consist of youth pedestrian injury counts in 603 census tracts of the New York City boroughs Bronx and Manhattan in the period from 2005 to 2014. The considered injury data were originally compiled by \citet{morris2019bayesian} and contain census tract level information about reported youth pedestrian injury counts (aggregated across different levels of injury severity), social fragmentation, traffic volume and private vehicle commute mode shares. The youth pedestrian injury counts are informed by the number of 5- to 18-year-old pedestrian injured in traffic crashes. Social fragmentation is measured by a composite index which takes into account the number of vacant housing units, single-person households, non-owner occupied housing units, and the population having relocated within the past year. Traffic volume is measured in terms of the maximum annual average daily traffic in the census tract. For more information about the data compilation and the data sources, the reader is directed to \citet{morris2019bayesian}. We supplement the data collected by \citet{morris2019bayesian} with information about the employment density (number of workers per km\textsuperscript{2}), the proportion of households with poverty status and the proportion of the population that identifies as Black or African-American. The supplementary data were sourced from the 2012--2016 American Community Survey \citep{uscensus2020acs}. Summary statistics for the considered data are reported in Table \ref{tab:sample_description}. Figures \ref{fig:map_obs_counts} and \ref{fig:hist_obs_counts} visualise the distribution of observed youth pedestrian injury counts across census tracts. A 5-nearest neighbour matrix for the study area is constructed using the PySAL library \citep{rey2010pysal} for Python. 

\begin{table}[H]
\centering
\small
\begin{tabular}{lrrrr}
\toprule
\textbf{Variable} &  \textbf{Mean} &  \textbf{Std.} &  \textbf{Min.} &   \textbf{Max.} \\
\midrule
Youth pedestrian injury count, 2005-14                       &  9.69 &  8.35 &  0.00 &  44.00 \\
Prop. of households with poverty status, 2012-16             &  0.24 &  0.15 &  0.00 &   0.57 \\
Prop. of black or African-American alone population, 2012-16 &  0.24 &  0.22 &  0.00 &   0.91 \\
No. of workers per km\textsuperscript{2} in 1000, 2012-16    & 17.96 & 37.34 &  0.02 & 260.40 \\
Social fragmentation index                                   &  2.02 &  2.73 & -4.50 &  18.67 \\
Avg. annual daily traffic (AADT) in 10k, 2015                &  4.45 &  4.68 &  0.21 &  27.65 \\
Private vehicle commute mode share, 2010-14                  &  0.19 &  0.15 &  0.00 &   0.76 \\
\bottomrule
\end{tabular}

\caption{Description of youth pedestrian injury counts and explanatory variables by census tract (N = 603)} \label{tab:sample_description}
\end{table}

\subsection{Goodness of fit} \label{sec:accu}

We evaluate the estimation accuracy of the MCMC and INFVB estimators in terms of goodness of fit to the training data. To this end, we compute three proper scoring rules, namely the log-score, the Dawid-Sebastiani score and the ranked probability score. In principle, a scoring rule provides a measurement of the discrepancy between the observed outcome and the estimated predictive distribution. A scoring rule is said to be proper if the expected score is minimised by the true predictive distribution \citep{gneiting2007strictly, wei2014calibration}. The three considered scoring rules are defined and calculated as follows:
\begin{itemize}
\item The log-score \citep[LS;][]{gneiting2007strictly, wei2014calibration} corresponds to the negative pointwise log-likelihood:
\begin{equation}
\text{LS}(y_{\text{obs}}, \theta) = - \log f(y_{\text{obs}} \vert \theta).
\end{equation}
For the NB model, the log-score is given by 
\begin{equation}
\text{LS}(y_i, \psi_i, r) = - \ln \Gamma(y_i + r) + \ln \Gamma(r) + \ln \Gamma(y_i + 1) - y_i \psi_i + (y_i + r) \ln \left( 1 + \exp (\psi_i) \right).
\end{equation}
\item The Dawid-Sebastiani score \citep[DSS;][]{dawid1999coherent} is informed by the mean $\mu$ and the variance $\sigma^2$ of the predictive distribution:
\begin{equation}
\text{DSS}(y_{\text{obs}}, \mu, \sigma^2) = \frac{(y_{\text{obs}} - \mu)^2}{\sigma^2} + \log \sigma^{2}.
\end{equation}
For the NB model, we have $\mu_i = \exp (\psi_i) r$ and $\sigma_{i}^{2} = \left( \exp (\psi_i) + \exp (2 \psi_i) \right ) r$.
\item The ranked probability score \citep[RPS;][]{matheson1976scoring} depends on the whole predictive distribution:
\begin{equation}
\text{RPS}(F, y_{\text{obs}}) = \sum_{t=0}^{\infty} \left( F(t) - \mathbb{1}\{y_{\text{obs}} \leq t \} \right)^{2},
\end{equation}
where $F$ denotes the predictive cumulative distribution function (CDF). $\mathbb{1}\{y_{\text{obs}} \leq t \}$ is an indicator which is one if the observed outcome $y_{\text{obs}}$ is less than the threshold $t$ and zero otherwise.
\citet{jordan2019evaluating} and \citet{wei2014calibration} provide expressions for the ranked probability score of the NB model:
\begin{equation}
\begin{split}
\text{RPS}(F_{r, p_i}, y_i) = & y_i \left ( 2 F_{r, p_i}(y_i) - 1 \right ) - \frac{r p_i}{(1 - p_i)^{2}} \\
& \left ( (1 - p_i) \left ( 2 F_{r+1, p_i} (y_i - 1) - 1 \right ) + _{2}\mathcal{F}_{1}\left( r + 1, \frac{1}{2}; 2; - \frac{4 p_i}{(1 - p_i)^{2}} \right) \right ).
\end{split}
\end{equation}
Here, 
$F_{r, p}(y) = 
\begin{cases} 
1 - I_p (y+1, r), & y \geq 0 \\
0 & y < 0 
\end{cases}
$
is the CDF of the NB distribution; $I_x (a, b)$ represents the regularised incomplete beta function; $_{2}\mathcal{F}_{1}(a, b; c; z)$ denotes the hypergeometric function.
\end{itemize}
For simplicity, the definitions presented above pertain to a single observation. In practice, aggregate scores are computed by summing over all observations in the data. In a Bayesian context, the posterior distributions of the scores can be obtained by evaluating the scores at the posterior samples of the model parameters.

\subsection{Results}
For the case study, the same estimation practicalities as for the simulation study (see Section \ref{sec:imp_est_prac}) apply with the only a minor difference that for INFVB, the grid over $\tau$ consists of 16 equidistant points in the interval $[-1.5,0]$.

Our first finding is that INFVB is substantially faster than MCMC. While the estimation time of MCMC is 135.9 minutes, the estimation of INFVB is only 2.9 minutes. The computation time of INFVB can be further decreased by distributing step 1 of Algorithm \ref{alg:infvb} over more than eight computer cores. In theory, as many compute cores as there are grid points can be used and the estimation time of INFVB can be further decreased by a factor of 20. However, it is important to note that the MCMC simulation cannot be sped further due to the sequential and conditional nature of Gibbs sampling.

The goodness of fit results of the MCMC and INFVB estimators are compared in Table \ref{tab:gof_real}. For all scores, the posterior mean of INFVB is marginally smaller than the respective posterior mean of MCMC. For example, the posterior mean of the Dawid-Sebastiani score for MCMC is 2762.3, while it is 2720.2 for INFVB. For all scores, the credible intervals of MCMC are wider than those of INFVB. In fact, the credible intervals of the INFVB scores are fully contained within the MCMC credible intervals. In a nutshell, the posterior distributions of the scores indicate that MCMC and INFVB provide the same level of goodness of fit to the training data, while MCMC estimation carries greater uncertainty than INFVB estimation. Lower uncertainty in INFVB estimates is as expected and is consistent with the literature \citep{blei2017variational,giordano2018covariances}. 

\begin{table}[H]
\centering
\small
\begin{tabular}{l | lll | lll}
\toprule
{} & \multicolumn{3}{c|}{\textbf{MCMC}} & \multicolumn{3}{c}{\textbf{INFVB}} \\
\textbf{Score} &    \textbf{Mean} &    \textbf{[2.5\%;} &  \textbf{97.5\%]} &    \textbf{Mean} &    \textbf{[2.5\%;} &  \textbf{97.5\%]} \\
\midrule
LS  &  1846.3 &  [1785.2; &  1878.1] &  1832.5 &  [1770.8; &  1855.7] \\
DSS &  2762.3 &  [2588.0; &  2864.7] &  2720.2 &  [2552.0; &  2796.3] \\
RPS &  2159.6 &  [1953.9; &  2275.4] &  2102.5 &  [1858.2; &  2192.0] \\
\bottomrule
\end{tabular}

\caption{Goodness of fit to youth pedestrian injury count data by estimation method} \label{tab:gof_real}
\end{table}

\begin{figure}[H]
\centering
\includegraphics[width = \textwidth]{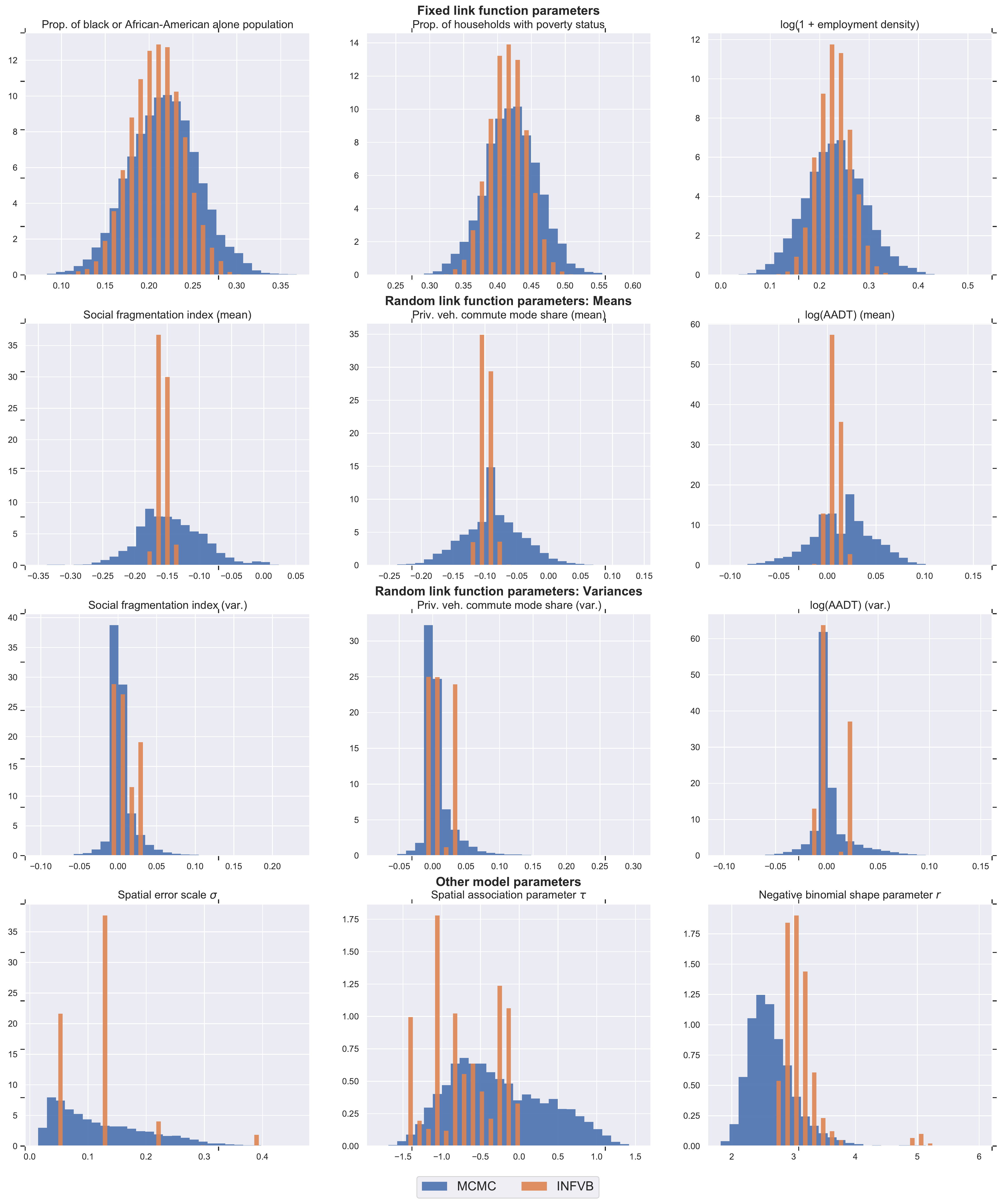}
\caption{Marginal posterior approximations of MCMC and INFVB for the youth pedestrian injury count data} \label{fig:posterior_real}
\end{figure}

Figure \ref{fig:posterior_real} shows the marginal posterior approximations inferred by MCMC and INFVB of selected model parameters. By and large, the posterior approximations produced by the two methods exhibit a close correspondence. In particular, the posterior approximations of the fixed link function parameters, the mean and variance terms of the random link function parameters, the spatial error scale $\sigma$ and the spatial association parameter $\tau$ coincide closely. For the the negative binomial shape parameter $r$, the posterior approximations of MCMC and INFVB overlap, but their modes differ.

\begin{figure}[H]
\centering
\includegraphics[width = 0.7 \textwidth]{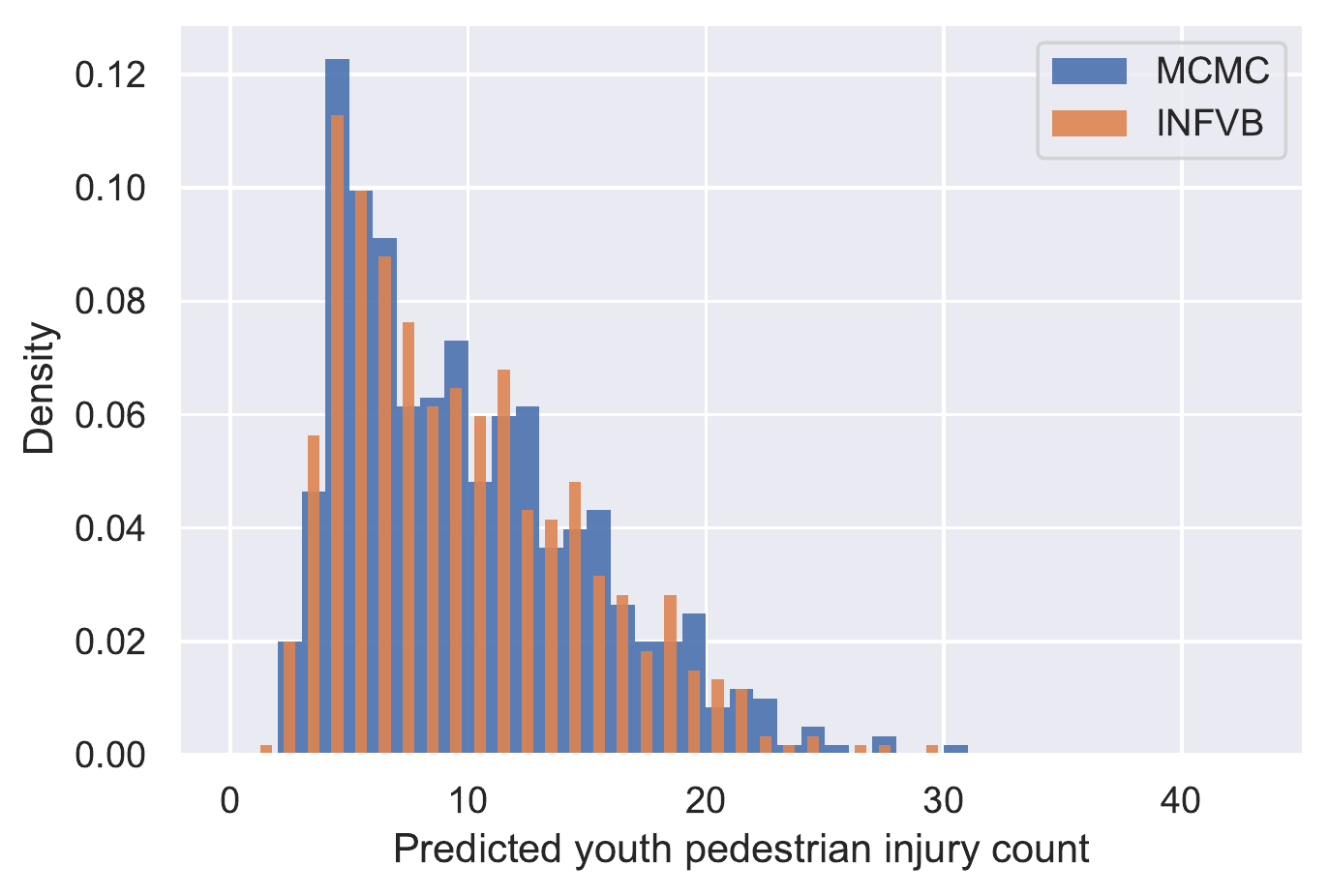}
\caption{Histogram of predicted youth pedestrian injury counts in the Bronx and Manhattan by census tract and estimation method} \label{fig:hist_prediction}
\end{figure}

\begin{figure}[H]
\centering
\includegraphics[width = 0.8 \textwidth]{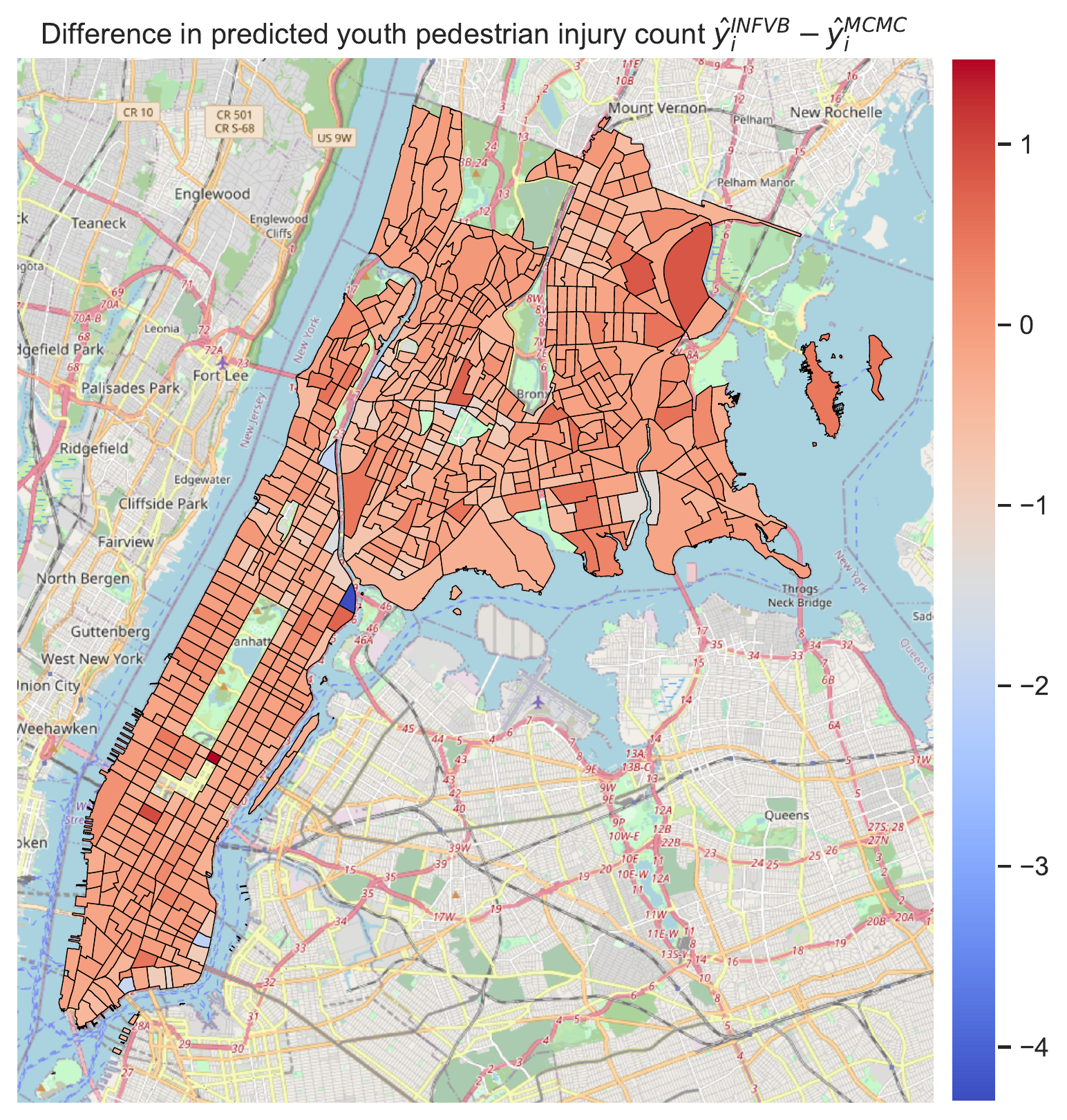}
\caption{Differences in youth pedestrian injury counts predicted by INFVB and MCMC in the Bronx and Manhattan by census tract} \label{fig:map_diff_prediction}
\end{figure}

Furthermore, we contrast the in-sample predictive accuracy of the MCMC and INFVB estimators by comparing the predicted injury counts for each census tract. Figure \ref{fig:hist_prediction} shows histograms of the predicted injury counts for both MCMC and INFVB. It can be seen that the two distributions overlap closely with each other. In addition, Figure \ref{fig:map_diff_prediction} visualises the difference between the youth pedestrian injury counts predicted by INFVB ($\hat{y}^{\text{INFVB}}$) and the corresponding MCMC prediction ($\hat{y}^{\text{MCMC}}$) for all census tracts. The differences in predicted youth pedestrian injury counts are generally small relative to the observed injury counts (see Figure \ref{fig:map_obs_counts}).
 
\begin{figure}[H]
\centering
\includegraphics[width = 0.7 \textwidth]{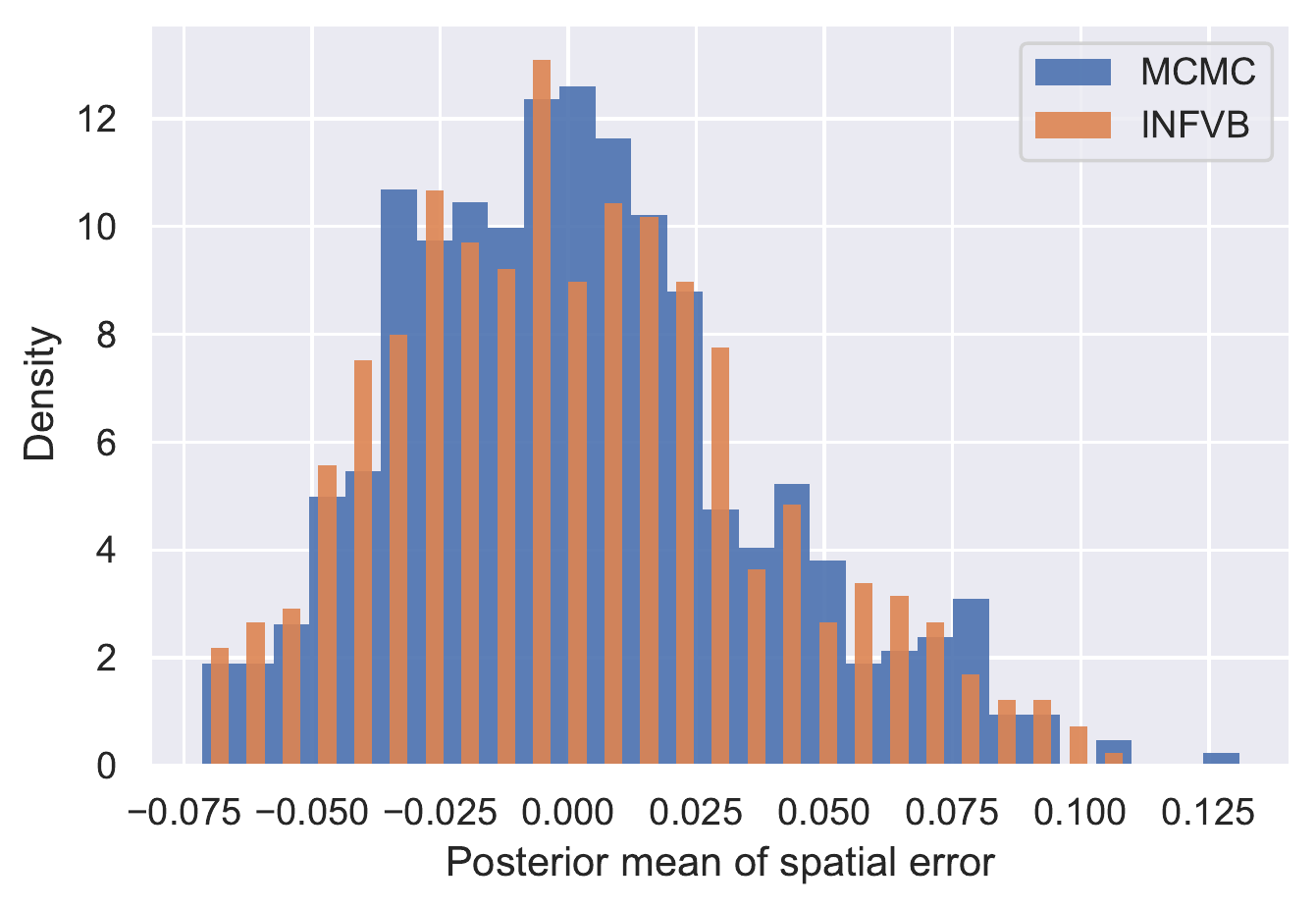}
\caption{Histogram of posterior means of spatial errors $\{\phi_{1}, \ldots, \phi_{N} \}$ by census tract and estimation method} \label{fig:hist_spatial}
\end{figure}

Finally, Figure \ref{fig:hist_spatial} shows histograms of the posterior means of the spatial errors $\{\phi_{1}, \ldots, \phi_{N} \}$ for MCMC and INFVB. The figure suggests that MCMC and INFVB perform equally well at recovering the unobserved spatial dependence.

\section{Conclusion}  \label{sec:conc}  

In this paper, we propose and empirically validate a variational Bayes (VB) method for posterior inference in a negative binomial model with unobserved spatial heterogeneity and dependence. The proposed VB method relies on P{\'o}lya-Gamma data augmentation to deal with the non-conjugacy of the negative binomial likelihood and an integrated non-factorised specification of the variational distribution to capture posterior dependencies. We benchmark the proposed VB method against MCMC using simulated data as well as real data on youth pedestrian injury counts in the census tracts of the New York City boroughs Bronx and Manhattan. In both applications, the VB approach is around 45 to 50 times faster than MCMC on a regular eight-core processor and emulates the estimation and predictive accuracy of MCMC. The marginal posterior approximations inferred by the VB approach and MCMC also resemble each other closely. The sequential and conditional nature of Gibbs sampling precludes improvement in computational efficiency through parallelisation. By contrast, INFVB can be further accelerated by a factor of up to 20 by taking full advantage of its embarrassingly parallel nature. Thus, INFVB is a scalable alternative to MCMC for the estimation of spatial count data models.    

There are several ways in which future work can extend the research presented in the current paper. First, MCMC and VB should be compared on other data sets from other disciplines to collect additional evidence about the relative advantages of the two methods. A second directions for future work is to adapt the proposed VB approach to models with spatio-temporal dependencies. Finally, recent advances in stochastic optimisation could be leveraged to enable the application of the proposed VB method to online inference problems \citep{hoffman2013stochastic}. Online estimation updates parameters continually, as new data points arrive, and thus facilitates the processing of very large data sets and data streams. 

\section*{Acknowledgements}

We would like to thank the associate editor and two anonymous reviewers for their critical assessment of our work.
Furthermore, we are grateful to Michel Bierlaire for his helpful comments and suggestions. 

\section*{Author contribution statement}

PB: conception and design, method derivation, manuscript writing and editing.
RK: conception and design, method implementation, data preparation and analysis, manuscript writing and editing. 
DJG: resources, manuscript editing. 

\newpage
\bibliographystyle{apalike}
\bibliography{bibliography.bib}

\newpage
\begin{appendices}

\section{Conditional posterior update of \texorpdfstring{$r$ in MCMC}{r}}\label{app:r_mcmc} 

To obtain the conditional posterior distribution of the dispersion parameter $r$ in MCMC, we follow the strategy adopted by \cite{zhou2012lognormal}. We represent the negative-binomial-distributed count variable as follows: 
\begin{equation*}
    y_{i} = \sum_{i=1}^{L_{i}} \chi_{li}, \quad \quad L_{i} \sim \text{Poisson}(-r\ln(1-p_{i})),  \quad \quad \chi_{il} \overset{iid}{\sim} \text{Logarithmic}(p_{i}).
\end{equation*}

Thus, the conditional posterior update of $r$ is: 
\begin{equation} \label{eq:r_mcmc}
\begin{split}
    P(r \lvert -) & \propto \prod_{i=1}^{N} P(L_{i} \lvert r, p_{i}) P(r \lvert r_{0}, h), \\ r \lvert - & \sim \text{Gamma}\left(r_{0} + \sum_{i=1}^{N} L_{i}, h+ \sum_{i=1}^{N} \ln(1+\exp({\psi}_{i})) \right). \\
\end{split}    
\end{equation}

Since the posterior update of $r$ is conditional on $\bm{L}$, we also update the conditional posterior of $L_i$ using the following equation:
\begin{equation}\label{eq:L_mcmc}
    P(L_{i} = j \lvert -)  = R(y_{i},j) \quad  j=\{0,1,\dots,y_{i}\} ,
\end{equation}
\[ 
R(l,m)= \left\{
\begin{array}{ll}
      1 & l=0; m=0 \\
      \frac{F(l,m) r^{m}}{\sum_{j=1}^l F(l,j) r^{j}} & l \neq 0; m \neq 0 ,\\
\end{array} 
\right. 
\]
\[ 
F(m,j)= \left\{
\begin{array}{ll}
      1 & m=1 \; \& \; j=1 \\
      0 & m < j \\
      \frac{m-1}{m}F(m-1,j) + \frac{1}{m}F(m-1,j-1) & 1 \leq j \leq m.\\
\end{array} 
\right. 
\]

\section{Supplementary material for INFVB}

\subsection{Important expressions and identities} \label{app:imp_id}

\begin{alignat*}{3}
\mathbb{E}[\bm{\Omega}] & = \begin{bmatrix} 
   \mathbb{E}[\omega_{1}] & \dots & 0 \\
    \vdots & \ddots &  \vdots\\
    0 &    \dots    &  \mathbb{E}[\omega_{N}] 
    \end{bmatrix}_{N \times N},\;   \quad \quad \quad && \mathbb{E}[\omega_{i}] = \left(y_{i} + \frac{\Tilde{b}_{r}}{\Tilde{c}_{r}}\right) \mathbb{E}\left[\frac{\text{tanh} \left(\frac{{\psi}_{i}}{2}\right)}{2\psi_{i}} \right],  \\
    \mathbb{E}(L_{i}) & = \sum_{j=1}^{y_{i}} R_{\tilde{r}}(y_{i},j) j,\; \quad \quad \quad && \tilde{r} = \exp \left(\Psi(\Tilde{b}_{r}) - \log(\Tilde{c}_{r}) \right), \\
    \mathbb{E}[\bm{Z}^{*}] & = 
    \begin{bmatrix} 
    \mathbb{E}[Z_{1}^{*}]\\
   \vdots\\
    \mathbb{E}[Z_{N}^{*}]\\
    \end{bmatrix}_{N \times 1} = 
    \begin{bmatrix} 
    \frac{y_{1}-\frac{\tilde{b}_{r}}{\tilde{c}_{r}}}{2}\\
   \vdots\\
    \frac{y_{N}-\frac{\tilde{b}_{r}}{\tilde{c}_{r}}}{2}\\
    \end{bmatrix}_{N \times 1},\; \quad \quad \quad && \bm{\Lambda}_{\bm{\beta}}  =    \begin{bmatrix} 
   \bm{\Lambda}_{\bm{\beta}_{1}}  & \dots & 0 \\
    \vdots & \ddots &  \vdots\\
    0 &    \dots    &  \bm{\Lambda}_{\bm{\beta}_{N}}
    \end{bmatrix}_{NK \times NK}, 
\end{alignat*}
where $\Psi(.)$ is a digamma function. $\mathbb{E}\left[\log(1+\exp(\psi_{i}))\right]$ and $\mathbb{E}\left[\frac{\text{tanh} \left(\frac{{\psi}_{i}}{2}\right)}{2\psi_{i}} \right]$ are obtained using Gauss-Hermite quadrature \citep{abramowitz1948handbook}.

\subsection{Important expressions to update \texorpdfstring{$q^{*}(\bm{\Theta}^{(g)}_{d})$ }{theta^d}} \label{app:q_d}

\begin{equation} \label{eq:theta_d_upd}
\begin{split}
    \mathbb{E}\left[\ln q(\bm{\Theta}^{(g)}_{c} \lvert \bm{\Theta}^{(g)}_{d}) \right] & =  -\frac{1}{2} \ln \lvert \bm{\Lambda}_{\bm{\phi}}^{(g)} \lvert - \frac{1}{2} \ln \lvert \bm{\Lambda}_{\bm{\gamma}}^{(g)} \lvert - \sum_{i=1}^N \frac{1}{2} \ln \lvert \bm{\Lambda}_{\bm{\beta_{i}}}^{(g)} \lvert  - \frac{1}{2} \ln \lvert \bm{\Lambda}_{\bm{\mu}}^{(g)} \lvert + \sum_{k=1}^K \ln \Tilde{c}_{a_{k}}^{(g)} \\
    & -\frac{K+1}{2} \ln \lvert \Tilde{\bm{B}}^{(g)} \lvert + \ln \Tilde{c}_{h}^{(g)} - \Tilde{b}_{r}^{(g)} + \ln \Tilde{c}_{r}^{(g)} - \ln \Gamma(\Tilde{b}_{r}^{(g)}) - \left(1 - \Tilde{b}_{r}^{(g)}\right) \Psi\left(\Tilde{b}_{r}^{(g)}\right).  \\
    & \\
   \mathbb{E} \left[\ln P(\bm{y} , \bm{\Theta}^{(g)}_{c}, \bm{\Theta}^{(g)}_{d}) \right] & =
   \sum_{i=1}^{N}\left[\mathbb{E} \left[\ln \Gamma\left(y_{i} + r^{(g)}\right)\right] -\mathbb{E} \left[\ln \Gamma\left(r^{(g)}\right)\right] + y_{i} \lambda_{\psi_{i}}^{(g)} \right] \\
   & -\sum_{i=1}^{N}\left[\left( y_{i} + \left[\frac{\Tilde{b}_{r}}{\Tilde{c}_{r}}\right]^{(g)}\right)\mathbb{E} \left[ \ln \left(1 + \exp\left(\psi_{i}^{(g)}\right)\right)\right]\right]\\
   & +\frac{1}{2} \ln \lvert (\Tilde{\bm{\Omega}})^{(g)}\lvert - \frac{1}{2} \left( \left[{\bm{\lambda}_{\bm{\phi}}^{T}} \Tilde{\bm{\Omega}} \bm{\lambda}_{\bm{\phi}}\right]^{(g)} + \text{tr}\left(\bm{\Lambda}_{\bm{\phi}} (\Tilde{\bm{\Omega}}) \right)^{(g)}\right) -\frac{N}{2}\ln \lvert \Tilde{\bm{B}}^{(g)} \lvert   \\
   & -\frac{\tilde{\rho}}{2} \sum_{i=1}^N \left[ (\bm{\lambda}_{\bm{\beta}_{i}} - \bm{\lambda}_{\bm{\mu}})^{T}{{\Tilde{\bm{B}}}^{-1}} (\bm{\lambda}_{\bm{\beta}_{i}} - \bm{\lambda}_{\bm{\mu}}) + \text{tr}(\Tilde{\bm{B}}^{-1} \bm{\Lambda}_{\bm{\beta}_{i}}) +
   \text{tr}({\Tilde{\bm{B}}}^{-1} \bm{\Lambda}_{\bm{\mu}})
   \right]^{(g)} \\
   & + r_{0} \left(- \ln \Tilde{c}_h^{(g)}\right) + (r_0-1)\left(\Psi(\Tilde{b}_r^{(g)}) - \ln \Tilde{c}_r^{(g)}\right) -  \frac{\Tilde{b}_h\Tilde{b}_r^{(g)}}{\Tilde{c}_h^{(g)}\Tilde{c}_r^{(g)}} \\
   & + (1-b_{0}) \ln \Tilde{c}_h^{(g)} - c_{0}  \frac{\Tilde{b}_h}{\Tilde{c}_h^{(g)}} 
   -\frac{1}{2} (\bm{\lambda}_{\bm{\gamma}}^{(g)} - \bm{\zeta}_{\bm{\gamma}})^{T}\bm{\Delta}_{\bm{\gamma}}^{-1} (\bm{\lambda}_{\bm{\gamma}}^{(g)} - \bm{\zeta}_{\bm{\gamma}}) \\
 &  -\frac{1}{2}\text{tr}(\bm{\Delta}_{\bm{\gamma}}^{-1} \bm{\Lambda}_{\bm{\gamma}}^{(g)})  + (b_{\sigma^2} - 1) \ln \sigma_{(g)}^{-2} -  c_{\sigma^2}   \sigma_{(g)}^{-2} \\
   & -\frac{(\tau^{(g)} - \zeta_{\tau})^2}{2 \sigma_{\tau}^{2}} -\frac{1}{2} (\bm{\lambda}_{\bm{\mu}}^{(g)} - \bm{\zeta}_{\bm{\mu}})^{T}\bm{\Delta}_{\bm{\mu}}^{-1} (\bm{\lambda}_{\bm{\mu}}^{(g)} - \bm{\zeta}_{\bm{\mu}})  -\frac{1}{2}\text{tr}\left(\bm{\Delta}_{\bm{\mu}}^{-1} \bm{\Lambda}_{\bm{\mu}}^{(g)}\right) \\
   &+ \sum_{k=1}^{K} \left((1-s) \ln \Tilde{c}_{a_k}^{(g)} - \eta_{k}  \frac{\Tilde{b}_{a_k}}{\Tilde{c}_{a_k}^{(g)}} \right) \\
   & - \frac{\rho}{2}\sum_{k=1}^K \ln \Tilde{c}_{a_k}^{(g)}  -\frac{\rho+K+1}{2} \ln \lvert \Tilde{B}^{(g)} \lvert  - \nu \Tilde{\rho} \sum_{k=1}^K \frac{\Tilde{b}_{a_k}}{\Tilde{c}_{a_k}^{(g)}} \left(\Tilde{B}^{(g)}\right)_{kk}^{-1} .
    \end{split}
\end{equation}

Thus, the conditional ELBO of INFVB for the spatial negative binomial model is obtained by inserting expressions presented in equation \ref{eq:theta_d_upd} in the following equation:
\begin{equation} 
\text{Conditional ELBO} =   - \mathbb{E}_{q} \left [ \ln q\left(\bm{\Theta}^{(g)}_{c} \lvert \bm{\Theta}^{(g)}_{d}\right) \right ] + \mathbb{E}_{q} \left [ \ln P\left(\bm{y}, \bm{\Theta}^{(g)}_{c},\bm{\Theta}^{(g)}_{d}\right) \right ].
\end{equation}

The optimal conditional distribution of $\bm{\Theta}^{(g)}_{c}$ is obtained by maximising the conditional ELBO or equivalently minimising its negative at each grid point (as detailed in equation \ref{eq:invb_C}):  
\begin{equation} 
     q^{*}\left(\bm{\Theta}^{(g)}_{c} \lvert \bm{\Theta}^{(g)}_{d}\right) =  \operatorname*{arg\,min}_{q\left(\bm{\Theta}^{(g)}_{c} \lvert \bm{\Theta}^{(g)}_{d}\right)}  \mathbb{E}_{q} \left [ \ln q\left(\bm{\Theta}^{(g)}_{c} \lvert \bm{\Theta}^{(g)}_{d}\right) \right ] - \mathbb{E}_{q} \left [ \ln P\left(\bm{y}, \bm{\Theta}^{(g)}_{c},\bm{\Theta}^{(g)}_{d}\right) \right ].
\end{equation}
\end{appendices}

\end{document}